\documentclass[fleqn,usenatbib]{mnras}

\usepackage[T1]{fontenc}
\usepackage{graphicx}	
\usepackage{amsmath}	
\usepackage{amssymb}	
\usepackage{pdflscape} 

\usepackage[dvipsnames]{xcolor}
\usepackage{siunitx}
\usepackage{newtxtext,newtxmath}

\usepackage{tabularx}

\usepackage{booktabs}
\usepackage{amsmath}
\usepackage{siunitx}
\usepackage{titlesec}

\usepackage{etoolbox}

\usepackage{xcolor}

\newcommand{\angstrom}{\textup{\AA}}

\newcommand{\ergs}{erg\,s$^{-1}$}

\newcommand{\swift}{{\it Swift}}

\graphicspath{{./}{figures/}}

\graphicspath{{./}{figures/}}

\title[Continuum reverberation mapping of NGC\,4051]{The unusually red delay spectrum of the low-mass black hole AGN NGC\,4051 as revealed by intensive continuum reverberation mapping with the Las Cumbres Observatory}

\usepackage{orcidlink}
\author[M. Marculewicz et al.]{Marcin Marculewicz\orcidlink{0000-0002-1380-1785},$^{1}$\thanks{E-mail: mmarculewicz@wayne.edu}
Juan V. Hernández Santisteban\orcidlink{[0000-0002-6733-5556]},$^{2}$\thanks{E-mail: jvhs1@st-andrews.ac.uk}
Keith Horne\orcidlink{0000-0003-1728-0304},$^{2}$
Edward M. Cackett\orcidlink{0000-0002-8294-9281},$^{1}$
\newauthor
Hermine Landt,$^{3}$
Jonathan Gelbord\orcidlink{0000-0001-9092-8619},$^{4}$
Hartmut Winkler\orcidlink{0000-0003-2662-0526},$^{5}$
Marianne Vestergaard\orcidlink{0000-0001-9191-9837},$^{6, 7}$,
Aaron J. Barth\orcidlink{0000-0002-3026-0562},$^{8}$
\newauthor
Michael Goad,$^{9}$
Shai Kaspi\orcidlink{0000-0002-9925-534X},$^{10}$
Paulina Lira,$^{11}$
Christopher A.  Onken,$^{12}$ 
Diego~H. Gonz\'alez-Buitrago$^{13}$
\newauthor
and Stefano Valenti$^{14}$
\\
$^{1}$Department of Physics and Astronomy, Wayne State University, 666 W. Hancock St, Detroit, MI, 48201, USA\\
$^{2}$SUPA School of Physics and Astronomy, University of St Andrews, North Haugh, St Andrews KY16 9SS, Scotland, UK\\
$^{3}$Centre for Extragalactic Astronomy, Department of Physics, Durham University, South Road, Durham, DH1 3LE, UK \\
$^{4}$Spectral Sciences Inc., 30 Fourth Ave. Suite 2, Burlington, MA 01803, USA\\
$^{5}$Dept. Physics, University of Johannesburg, PO Box 524, Auckland Park 2006, Johannesburg, South Africa\\
$^{6}$Steward Observatory, University of Arizona, 933 North Cherry Avenue, Tucson, AZ 85721, USA\\
$^{7}$DARK, The Niels Bohr Institute, University of Copenhagen, Jagtvej 155, DK-2200 Copenhagen N, Denmark\\
$^{8}$ Department of Physics and Astronomy, 4129 Frederick Reines Hall, University of California, Irvine, CA 92697-4575, USA\\
$^{9}$ School of Physics and Astronomy, University of Leicester, University Road, Leicester, LE1 7RH, UK\\
$^{10}$School of Physics and Astronomy and Wise Observatory, Tel-Aviv University, Tel-Aviv 6997801, Israel\\
$^{11}$Department of Astronomy, Faculty of Physical and Mathematical Science, University of Chile, Santiago, Chile\\
$^{12}$Research School of Astronomy \& Astrophysics, The Australian National University, Canberra, ACT 2611, Australia\\
$^{13}$Instituto de Astronomía, Universidad Nacional Autónoma de México, Km 103 Carretera Tijuana-Ensenada, 22860 Ensenada B.C., México\\
$^{14}$Department of Physics, University of California, 1 Shields Avenue, Davis, CA 95616-5270, USA\\
}

\date{Accepted XXX. Received YYY; in original form ZZZ}

\begin{document}
\maketitle

\begin{abstract}

We present a two-year optical reverberation mapping campaign of NGC 4051, an active galactic nucleus (AGN) hosting a low-mass black hole ($8\times10^5$ M$_\odot$), using daily observations in seven photometric bands from Las Cumbres Observatory  augmented by archival data from Swift XRT and UVOT.
The light curves show correlated variability with wavelength-dependent lags broadly consistent with the standard accretion disc scaling, $\tau \propto \lambda^{4/3}$, and a pronounced $u$-band excess. However, the $i$ and $z_s$ lags are significantly larger than expected and cannot be explained by a combination of disc emission and diffuse continuum (DC) from the broad-line region (BLR), 
making NGC 4051 a notable lag-luminosity outlier.
The spectral energy distribution (SED) of the variable AGN component is markedly redder than the canonical accretion disc prediction, $F_\nu \propto \nu^{1/3}$, typically observed in more massive systems. We explore two scenarios to account for the red UV–optical SED and the anomalously large $i$ and $z_s$ lags: (a) SMC-like dust reddening ($E(B-V)\sim0.18$) combined with optically thick emission from the inner edge of the dusty torus; and (b) a dominant diffuse continuum contribution. We discuss the implications of each scenario within a comprehensive multi-wavelength framework.
\end{abstract}




\section{Introduction} \label{sec:intro}

Every massive galaxy and active galactic nucleus (AGN) hosts a supermassive black hole (SMBH) at its center \citep{KormendyHo2013}. 
Direct measurements of black hole mass and accretion disc size remain elusive with current observational techniques, as these regions are too compact to be spatially resolved using conventional imaging methods. Exceptions to this limitation are the measurements of M87$^*$ and Sgr\,A$^*$ via high spatial resolution interferometry observations \citep{EHT2019}, although this technique is both resource-intensive and currently applicable to just these two objects. {Moreover}, this is a wavelength regime in which one can only study radio plasma, not the gas that reveals information about the accretion flow/disc and the broad-line region. Therefore, direct imaging of these regions remains {unfeasible} using conventional methods that probe the emission from these components.

To investigate SMBH properties over a broader sample of AGNs, indirect methods are employed, such as reverberation mapping \citep[RM,][]{Blandford82,Kaspi2000,Peterson04,Cackett21RMreview}. Traditionally, RM has been used to infer the size of the broad-line region (BLR) by measuring time delays between the continuum emission and the response of the emission lines \citep{Blandford82,Peterson04,Shen2023_arixvonly}. Another important application of RM involves the time delays between the continuum variations at different wavelengths, particularly in the ultraviolet (UV) and optical bands \citep[e.g.,][]{Fausnaugh2016RM_NGC5548,Edelson19,JVHS_F9_2020}. These continuum delays are linked to the geometry, size, and temperature profile of the accretion disc, providing insights into the structure of the central engine. Previous studies have noted challenges in measuring these continuum delays due to the complex and non-variable response of the accretion disc \citep{Starkey2016}, contamination from host galaxy starlight, contamination of the disc delay signature by a more distant reprocessor e.g., the diffuse continuum from the BLR \citep{KoristaGoad2001, Korista2019}, and the difficulty in disentangling the effects of intrinsic AGN variability from external factors such as dust extinction or gravitational microlensing \citep{Morgan2010}. These issues highlight the need for intensive observing campaigns and refinement of lag measurement techniques. Campaigns with higher cadence monitoring, broader wavelength coverage, and longer baselines provide the necessary data quality to develop and test new methods for separating these contaminating effects and to better isolate disc reverberation signals.

Intensive broadband reverberation mapping (IBRM) campaigns are designed to overcome these challenges and improve our understanding of the SMBH accretion process. 
By monitoring the continuum emission across a wide range of wavelengths --  from the UV to the near-infrared (NIR) -- we obtain high-quality time delay measurements that 
determine the size and probe the radial structure of the accretion disc. 
The broadband approach can provide a more complete picture of the AGN's emission, as different wavelengths trace different regions of the accretion flow, revealing details about its geometry, temperature, and ionization structure \citep{Cackett2007reprocessing_model,Fausnaugh2016RM_NGC5548, JVHS_F9_2020}. 
Accurate measurements of continuum delays 
also help address some of the uncertainties in previous studies, such as the impact of anisotropic emission, variations in the disc's inner edge, and time-varying obscuration \citep{KawaguchiMori2011, Kara2021, MM2023}. 
Furthermore, IBRM campaigns that secure long-term monitoring of a large sample of AGNs offer the statistical power needed to probe relationships between SMBH mass, accretion rate, and host galaxy properties. By improving our understanding of the continuum variability and its connection to the underlying accretion disc, IBRM campaigns contribute to a more detailed and systematic exploration of SMBH growth across diverse AGN populations.

In this study, we analyse a bright and nearby narrow-line Seyfert-1 AGN -- NGC\,4051 
\citep[$z = 0.002336$,][]{2001_Verheijen_redshift}. This source hosts a relatively small black hole with a virial mass of $M_{\rm BH}=(7.8\pm2.3)\times10^5$\,M$_\odot$ \citep{BentzKatz2015} as measured from variations in the broad component of the H$\beta$ emission line. The lag of the H$\beta$ BLR is $\sim 2$~light-days \citep{Fausnaugh17}.
Correlated variations between the optical and NIR continuum emission were observed by \citet{Koshida2014}, who measured the inner radius of the dusty torus in this AGN to be approximately $\sim 11-24$~light-days. But, interestingly, the optical/NIR continuum emission in NGC\,4051 at rest-frame wavelengths $\sim 5000-10000$~\AA~is dominated by host galaxy light \citep{Landt2013}. NGC\,4051 is known to be strongly variable also at X-ray energies, often showing X-ray flaring behaviour \citep[e.g.,][]{Breedt2010, Kumari2024, Rani2025}.
These studies recover significant inter-band delays of $\sim1-2$~days between UV and optical, but reveal only a weak correlation between the UV--optical bands and X-rays. Furthermore, NGC\,4051 is one of a few local AGN where an independent distance has been measured via Cepheid variables \citep[$D_{\rm L}=16.6\pm0.3$~Mpc,][]{Yuan2021}, providing an opportunity to test AGN as a standard candle \citep[e.g.,][]{Cackett2007reprocessing_model}. 

The remainder of this paper is organised as follows: 
Section\,\ref{sec:obs} briefly describes the observations and data reduction of NGC\,4051. 
Section\,\ref{sec:results} discusses
the time series analysis, including
measurement of the time-delay spectrum, the trend of smaller/larger lags for faster/slower variations, and the fits deriving time delay distributions vs wavelength.
Section\,\ref{Sec:SED_flux} presents the flux-flux analysis used to decompose the light curves into separate SEDs for the constant and variable components of the AGN light.
Section\,\ref{Sec:Discussion} contains a discussion of our main results. We summarise our conclusions in Section\,\ref{sec:summary_conclusion}.

\section{Observations}
\label{sec:obs}

\subsection{Las Cumbres Observatory}
We observed NGC\,4051 with the Las Cumbres Observatory (LCO) global robotic telescope network \citep{Brown:2013} in 7 optical imaging filters ($u, B, g, V, r, i, z_s$) across two observing seasons, 2021\,Nov\,22 through 2022\,Jul\,07 (Year\,1) and 2022\,Nov\,04 through 2023\,Jul\,28 (Year\,2). The LCO network enabled a median cadence of $\sim0.98$ days, where all 7 filters were observed at every visit. In total, $\sim770$ individual measurements were performed per filter, after discarding outlier measurements (due to aspects such as weather, cosmic rays, and instrumental artifacts). Basic data reduction (bias and flat-field corrections) was performed by the LCO standard pipeline {\sc Banzai} \citep{mccully2018}. The data were processed and retrieved from the LCO science archive.\footnote{\url{https://archive.lco.global}} We performed aperture photometry with a 5$^{\prime\prime}$ radius centred at the AGN using our customized pipeline based on {\sc SExtractor} \citep{bertin:1996}. We refer to \citet{JVHS_F9_2020} for further details on the data extraction and spectro-photometric calibration.

\subsubsection{Inter-telescope calibration} \label{sec:intercalibration}
Despite their identical design, the LCO 1-m robotic telescopes have small differences in their filter bandpasses, causing telescope-specific systematic offsets in our AGN light curves \citep{Vieliute2025}.
We follow the procedure from \citet[][for details see their Section\,2]{Donnan23_PG1119}, which makes small telescope-dependent multiplicative and additive corrections to bring together the light curve segments measured by different LCO telescopes using {\sc PyROA}\footnote{\url{https://github.com/FergusDonnan/PyROA}} \citep{pyroa21,Donnan23_PG1119}. We applied the same procedure for all bands, producing the data shown in Fig.~\ref{fig:2year_fit}. 
In addition, the inter-calibration procedure implements a sigma-clipping algorithm whereby outliers with residuals larger than a $4\,\sigma$ threshold have their error bars expanded to keep the residual at $4\,\sigma$. This adjustment is made to account for underestimated uncertainties in a few measurements affected by variable observing conditions, and only a small number of data points were modified in this way.




However, we faced an additional issue around MJD$~\simeq59700$ (end of the first year), where the clipping procedure expanded the error bars of {one to two} data points at the edge of the seasonal gap, {as the light curve could not be well constrained in that interval}. {These points were subsequently} excluded from the cross-correlation function (CCF) analysis. Therefore, we did not use the affected data points in the lag measurement. For completeness, we visually inspected all bands to ensure that no clear outliers remained.

\subsection{\swift\ Observatory}
\label{sec:xray_spec}

The Neil Gehrels \swift\ Observatory \citep[\swift\ hereafter,][]{Gehrels:2004} observed NGC\,4051 twice during the intensive monitoring period with LCO, on 2022\,Mar\,27 and 2022\,Mar\,29. The UVOT \citep{UVOT2005} photometry of NGC\,4051 was obtained concurrently with the X-ray observations. 
{In addition, we retrieved archival data over the last decade in all six filters (three UV bands – UVW1, UVM2, UVW2 – and three optical bands – U, B, V) to measure} the AGN spectral changes via the flux-flux analysis presented in Sec.~\ref{Sec:SED_flux}. The UVOT light curves were extracted using a 5" radius aperture, and dropout measurements were excluded from the light curves using the {detector} sensitivity {masks} \citep[see details in][]{JVHS_F9_2020}. The UVOT photometric measurements were not sufficient to retrieve inter-band delays due to the small number of epochs. However, they served as a crucial reference bandpass at shorter wavelengths, 
facilitating the retrieval of the spectral energy distribution (SED) as described in Sec.~\ref{Sec:SED_flux}.

\section{Time Series Analysis} \label{sec:results}

We analysed the multi-band photometric LCO light curves to measure inter-band time lags with different methods.
Our primary method (Section\,\ref{sec:pyroa}) models the light curves with a running optimal average (ROA) as implemented in {\sc PyROA} \citep{pyroa21}.\footnote{We also measured the inter-band delays with the widely used interpolated cross-correlation method as implemented by the {\sc PyCCF} software \citep{PyCCF_MSun_2018}. Both measurements and their comparison to the {\sc PyROA} lag measurements with cross-correlation lags measured can be found in Appendix~\ref{sec:pyccf}.}
In Section\,\ref{sec:timescale_lags}, we investigated the timescale-dependent lags via a grid of ROA timescales to quantify the trend of smaller/larger lags arising from faster/slower variations. Finally, we present evidence for an asymmetric delay distribution for each of the LCO bands in Section\,\ref{sec:memecho} using MEMEcho \citep{Horne1991}.

\begin{figure*}
\centering
    \includegraphics[width=0.9\textwidth]{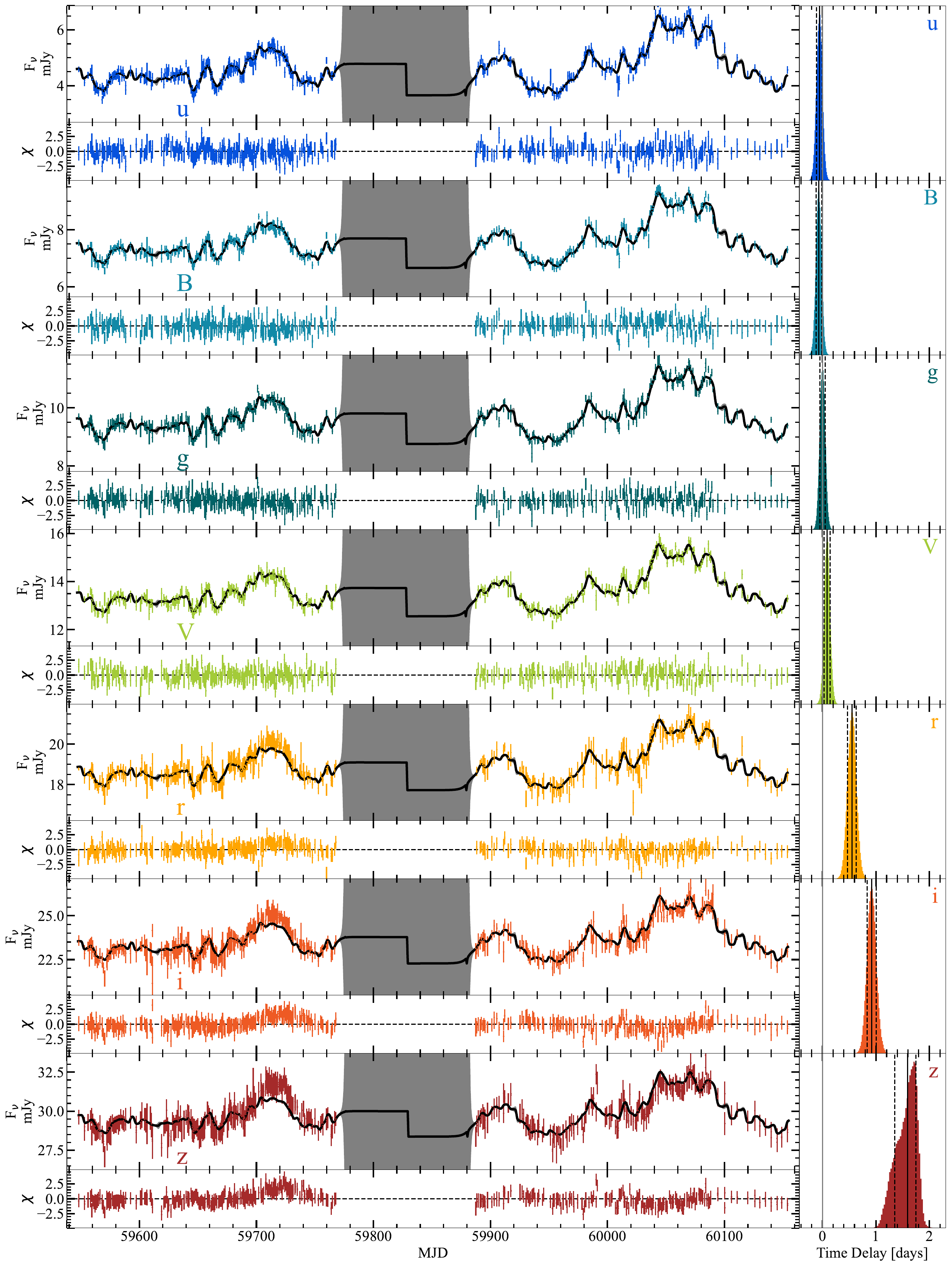}
    \caption{\label{fig:2year_fit}{\sc PyROA} fit to 7-band LCO light curves spanning 2 years. The ROA lightcurve shape $X(t)$ has a Gaussian smoothing width $\Delta=1.5$\,d. Each light curve panel shows data (colour-coded error bars) and the best-fit model light curve (solid black curve) with its $1\sigma$ uncertainty envelope (grey band), which expands greatly during the seasonal gap. All fluxes are in mJy. The subpanel below each light curve shows the normalized residuals ($-5<\chi<+5)$ to the jointly fitted {\sc PyROA} model.
    Note that systematic trends remain in the residuals for the $r$, $i$ and $z_s$ bands. The right panel shows the marginalized posterior distribution for the time lag $\tau$, relative to the $g$ band light curve, with the median value and 68\% confidence interval indicated by vertical solid and dashed black lines, respectively. Time delays are in observed frame days. The measured lags in each band are summarised in Table \ref{tab:timedelays-two_years_sim}.
    }
    \end{figure*}

\begin{figure*}
\centering
    \includegraphics[width=0.9\linewidth]{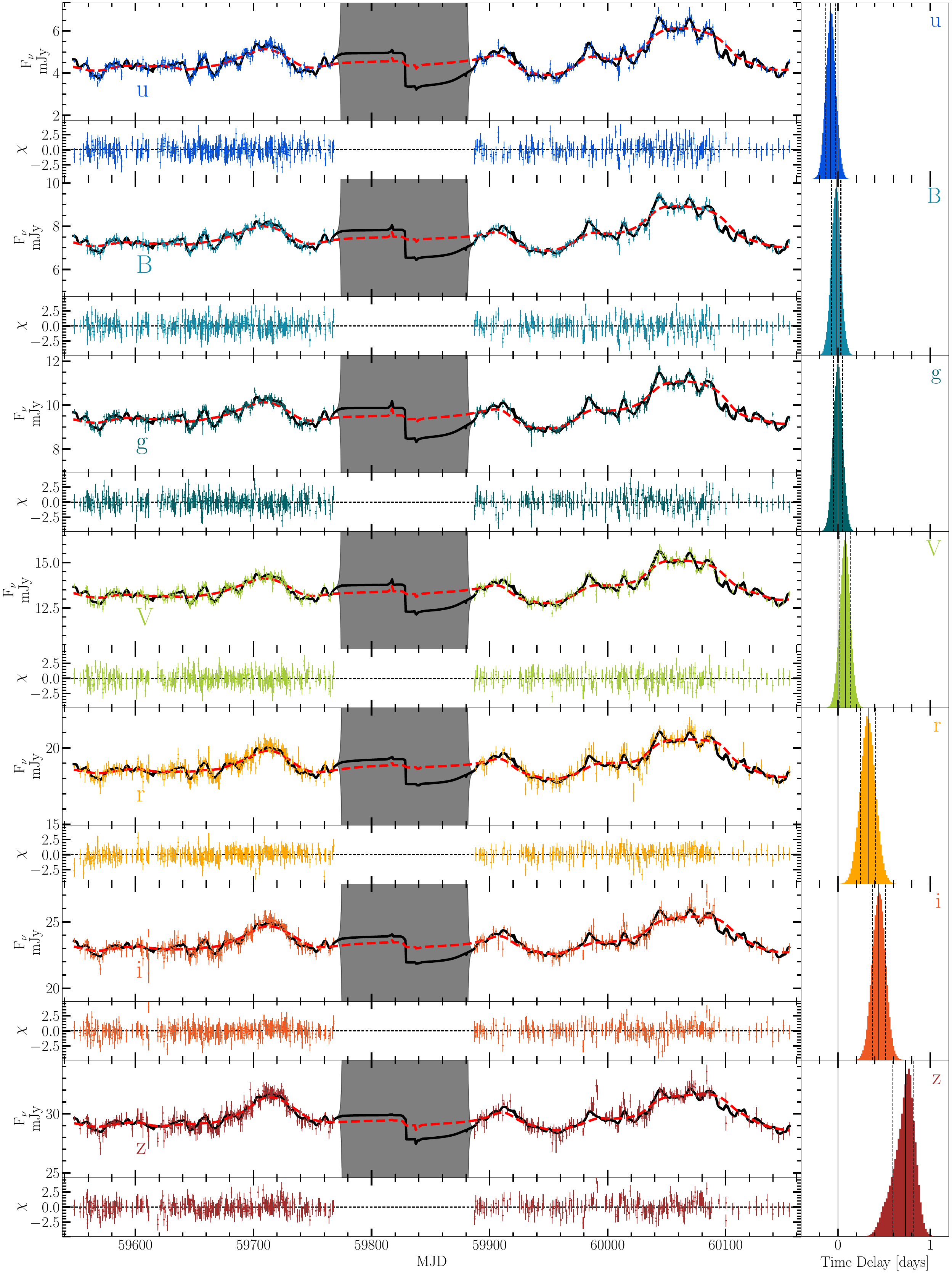}
    \caption{\label{fig:2year_wslow}As in Fig.\,\ref{fig:2year_fit} except that the LCO light curves are de-trended by subtracting the ROA model with $\Delta=10$\,d, (red dashed curves) and the ROA model with $\Delta=1.5$\,d {(black line)} is fitted to the residuals.
    Note that systematic trends in the residuals are now largely eliminated. The resulting lags are also smaller by a factor $\sim2$, indicating that variations removed by the 10~d detrending are delayed relative to those captured by the $\Delta=1.5$~d ROA model.
    }
\end{figure*}
\subsection{{\sc PyROA} analysis}\label{sec:pyroa}

We use the modelling tool {\sc PyROA} \citep[for details see][]{pyroa21, Donnan23_PG1119} to simultaneously fit the 7-band LCO light curves and determine their inter-band delays. 
{\sc PyROA} uses all light curves to define a dimensionless light curve shape, $X(t)$ -- normalised to zero mean, $\langle X(t)\rangle=0$, and unit variance, $\langle X(t)^2\rangle=1$. 
This is generated by an optimal (inverse-variance weighted) average over all the data, weighted by a Gaussian memory function with a width $\Delta$. 
The normalised lightcurve shape $X(t)$ is then scaled and shifted to fit each lightcurve with a different background level $\bar{F}(\lambda)$, variation amplitude $\Delta F(\lambda)$, and time shift $\tau(\lambda)$. 
Thus, the {\sc PyROA} model for each light curve at different wavelengths/filters is defined as:
\begin{equation}
     \label{eqn:pyroa}
    F(\lambda,t) = \bar{F}(\lambda) + \Delta F(\lambda)\, X\left( t - \tau(\lambda) \right)
    \ .
\end{equation}

The noise model used in the {\sc PyROA} fit includes, for each light curve, an additional variance $\sigma^2(\lambda)$ that is added in quadrature to the statistical uncertainties of the corresponding filter data. 
This accounts for possible systematic errors not included in the nominal error bars, and has the effect of down-weighting the influence of light curves with larger differences between the ROA model and the data.

{\sc PyROA} employs a Markov Chain Monte Carlo (MCMC) procedure \citep{emcee} to sample the posterior parameter distributions around the best-fit parameters that minimise the Bayesian Information Criterion \citep[BIC, see further details in][]{pyroa21}.
We ran the MCMC sampler until the chain length was more than 50 times longer than the autocorrelation length.
This ensures a robust convergence of the fit with enough effectively independent samples for parameter uncertainty estimates.

Fig.\,\ref{fig:2year_fit} shows a simultaneous {\sc PyROA} fit with $\Delta=1.5$\,days (hereinafter d) to the 7-band LCO light curves for the full 2-year dataset
\footnote{
We also ran {\sc PyROA} fits to Years\,1 and 2 separately. We discuss the resulting lag spectra with those found using {\sc PyCCF} in Appendix\,\ref{sec:pyccf}.
},
with lags relative to the $g$ band
light curve.
Our initial {\sc PyROA} fit,
minimising the BIC, finds an optimal solution with $\Delta=0.65\pm0.02$\,d,
with lags listed in the first column of Table\,\ref{tab:timedelays-two_years_sim}.
Such small values of $\Delta$ result in a very flexible $X(t)$ and rapid expansion of the ROA uncertainty envelope in data gaps as small as a few days. Although the BIC criterion favours these models, we find that they sometimes generate inappropriate rapid excursions in $X(t)$ to fit individual data points that are isolated from their neighbours.
In this case, we see an unrealistic jump
in the lag, from $<1$\,d to $>3$\,d between
the $i$ and $z_s$ bands.
We traced this to a feature in $X(t)$ moving from before to after a single $z_s$ data point near the edge of the data gap near MJD=59620.
To avoid such overfitting, we opted to limit the flexibility of the ROA light curve shape $X(t)$ by setting a lower bound on the prior, $\Delta>1.5$~d. This stiffer ROA model follows the rapid variations while being less affected by isolated data points.
The resulting fit is shown in Fig.~\ref{fig:2year_fit} with relevant parameters in Table~\ref{tab:timedelays-two_years_sim}. 

The {\sc PyROA} model with $\Delta=1.5$\,d fits all the light curves fairly well despite the assumption that all bands have the same lightcurve shape. In particular, the shorter-wavelength bands, which contain the sharpest features, are well reproduced.
However, at longer wavelengths ($i$, $r$, and $z_s$), the fit leaves clear systematic trends in the residuals.
The model is systematically below or above the data ($\sim$MJD 59720 and 60040, respectively). The fit responds to these larger residuals by increasing the additional variance for these bands.
These slowly-varying features in the residuals are not seen in the bluer bands (even at higher-SNR) and have amplitudes that increase with wavelength, suggesting that an additional slowly varying and relatively red component contributes to the overall variability.

\begin{figure}
\centering
    \includegraphics[width= 0.49 \textwidth]{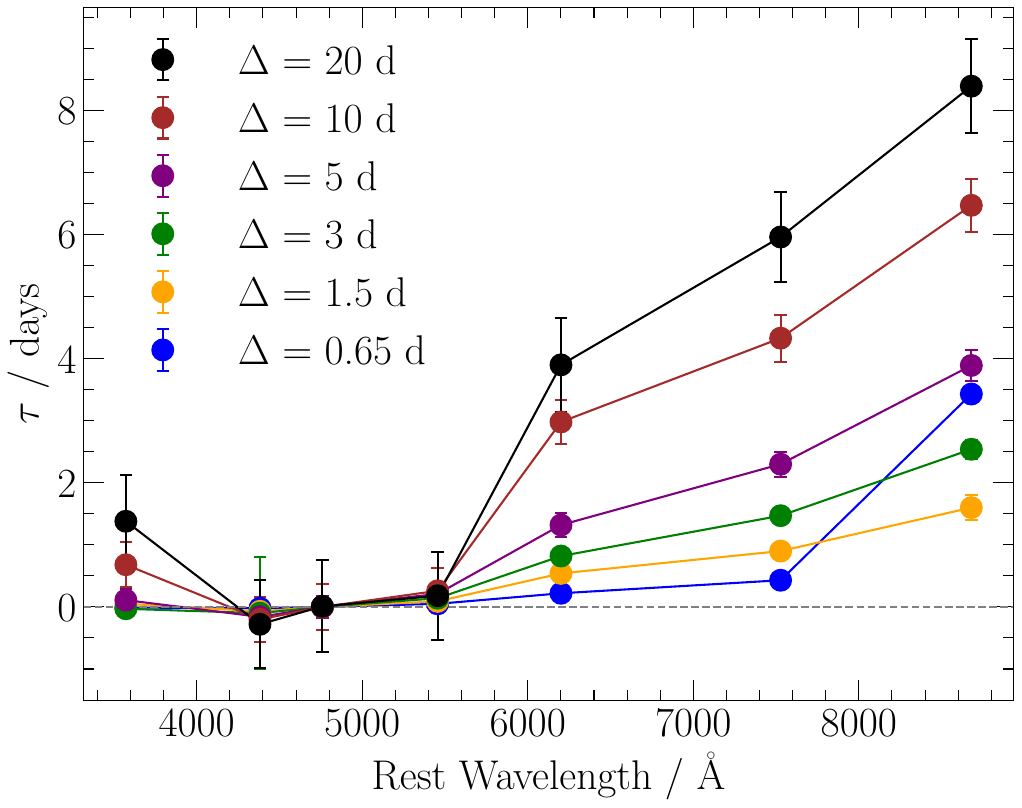}
    \includegraphics[width= 0.49 \textwidth]{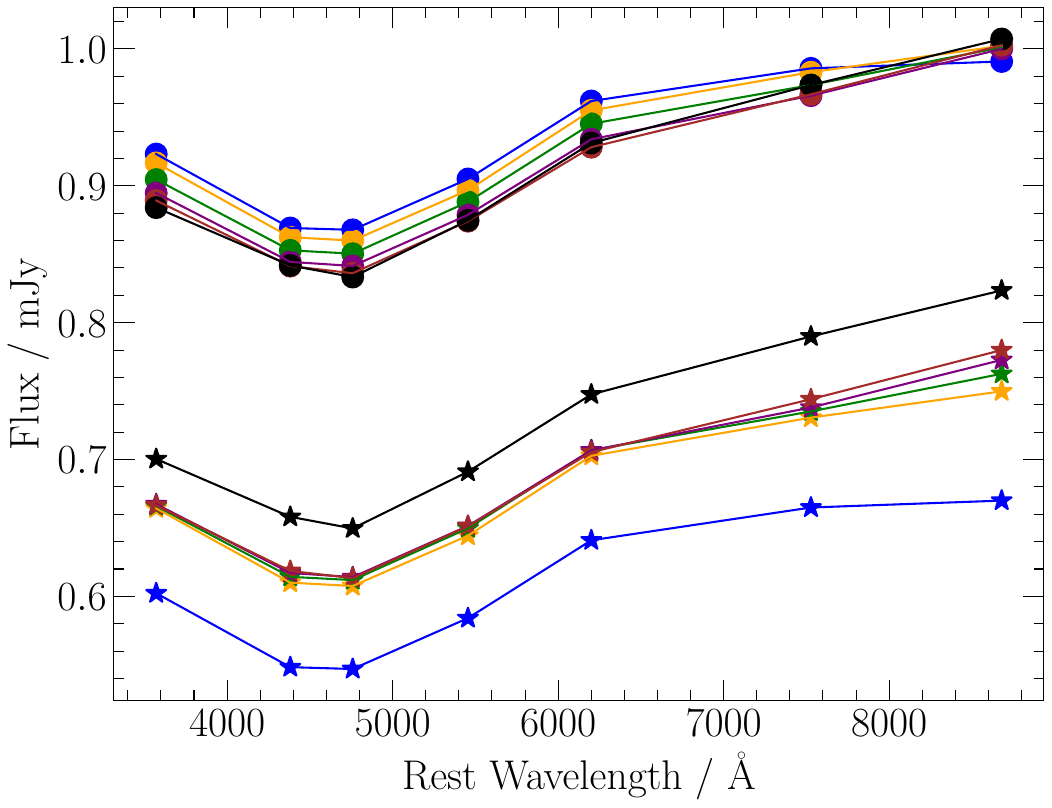}
    \caption{\label{fig:deltatrends}{\it Top:} The lag spectrum measured by {\sc PyROA} with different smoothing widths $\Delta$, showing the general trend for $\tau$ increasing with $\Delta$ at all wavelengths, but particularly in the $u$ band and the redder $r$, $i$, and $z$ bands. This pattern suggests a 2-component model with rapid response from the compact accretion disc plus free-bound continuum response from the more extended BLR. {\it Bottom:} The faint (marked as stars) and bright (marked as circles) disc SEDs measured from the mean and rms of light curves measured from {\sc PyROA} fits with different smoothing widths $\Delta$. 
    The spectral shape is independent of $\Delta$ and suggests bound-free emission with Balmer continuum sampled by the $u$-band and Paschen continuum sampled by the redder bands. The absolute scaling of the inferred bright disc SED is independent of $\Delta$ ($\Delta$ colors are the same as the top panel). The faint disc SED is lower for $\Delta=0.65$\,d and higher for $\Delta=20$\,d.}
\end{figure}

To improve the fit to these slowly-varying residual features in the longer-wavelength bands, we incorporate a slowly varying component into the {\sc PyROA} fit. 
{The smoothing timescale $\Delta$ controls the flexibility of the ROA: smaller values allow the ROA to track rapid variability but can result in spurious excursions when data points are isolated, while larger values enforce smoother behaviour at the cost of resolving short-timescale structure. Although the BIC-optimal value of $\Delta=0.65$\,d was found in the year-by-year fits (see Appendix), we find that this timescale produces anomalous behaviour in the $z$-band (Fig.\,\ref{fig:deltatrends}), likely because it is short enough to overfit isolated data points. We therefore adopt $\Delta=1.5$\,d as the timescale for the fast component, which captures the reverberation signal without these artefacts.} To model the slow variations, we fit all the data using {\sc PyROA}
with a smoothing timescale of $\Delta=10$\,d. {This value was chosen because it is long enough to suppress the dominant residual trends observed in the longer-wavelength bands, while remaining short enough that the SED does not vary significantly across the timescale; we found that larger values such as $\Delta=20$\,d begin to absorb genuine variability and distort the SED shape (Fig.\,\ref{fig:deltatrends}).}
We subtract this fit from the data and model the resulting residual light curves with a second {\sc PyROA} fit with $\Delta=1.5$\,d. 
Fig.\,\ref{fig:2year_wslow} shows the resulting fit where the red dashed curve in each panel shows the subtracted slow component. This model,
including slow variations with $\Delta=10$\,d and fast variations with $\Delta=1.5$\,d, more closely matches the observed light curves, reducing the residual trends shown in Fig.\,\ref{fig:2year_fit}.
The lags found have larger uncertainties and are smaller by a factor $\sim2$ when the 10~d variations are removed, suggesting that the 10~d variations are delayed relative to the faster variations captured by the $\Delta=1.5$\,d ROA model.
We investigate this trend of larger lags for slower variations below.

\subsubsection{{\sc PyCCF} vs {\sc PyROA}}

{
We also measured  inter-band lags with the widely used interpolated cross-correlation method as implemented by the {\sc PyCCF} software \citep{PyCCF_MSun_2018}. 
Appendix~\ref{sec:pyccflags} details
these measurements, made separately for Years\,1 and 2, and 
Appendix\,\ref{sec:yearbyyear_fits} 
compares the {\sc PyCCF} lags
with the corresponding {\sc PyROA} lags.
}

{ 
As shown in Fig.\,\ref{fig:ROAvsCCF_dividedby_years},
both {\sc PyROA} and {\sc PyCCF} reveal an overall trend of lags increasing with wavelength.
The CCF lags tend to fall in between the {\sc PyROA} lags for $\Delta=1.5$\,d and the longer lags for $\Delta=10$\,d.
This timescale-dependence of the lags likely arises from the asymmetry of the transfer function: the long positive-lag tail tilts the CCF near its peak, displacing the centroid to larger values. 
By contrast, {\sc PyROA} concentrates the lag estimates on variations near the ROA timescale $\Delta$.
The CCF lags are also larger in Year\,2 than in Year\,1, due to an increase in slower variations in the Year\,2 light curve.
In comparison, the {\sc PyROA} lags for both $\Delta=1.5$\,d and $\Delta=10$\,d are more consistent between the two years.
{\sc PyROA} also yields more precise lag measurements with uncertainties that are roughly five times smaller than those from {\sc PyCCF}. 
For these reasons, {\sc PyROA} is adopted as the primary analysis method.
}

\begin{table*}
    \centering
    \caption{\label{tab:timedelays-two_years_sim}Timescale dependent lags of NGC\,4051 as measured in the observed frame relative to the $g$-band light curve. These are measured with {\sc PyROA} on the two-year dataset simultaneously. All values of lags are in observed frame days. Values without uncertainties in the Gaussian width of the ROA were fixed in the timescale-dependent lag analysis, as described in Sec.~\ref{sec:timescale_lags}. 
    Final column gives the median delay from the MEMEcho maps shown in Fig. \,\ref{fig:Memecho}.}

\begin{tabular}{c|cccccc|c}
    \hline
    Filter & \multicolumn{6}{c}{ROA Timescale $\Delta$ (days)} & MEMEcho   \\
         \hline
     & $0.65\pm0.02$ & $1.5$ &$3$& $5$& $10$ & $20$ & median \\
    \hline       
    $u$ & $-0.01\pm0.03$ & $-0.05\pm0.05$ & $-0.03\pm0.10$ & $0.11\pm0.18$&$0.68\pm0.36$ &$1.38\pm0.76$ & $-0.07$\\
    $B$  & $-0.02\pm0.03$ & $-0.07\pm0.05$& $-0.10\pm0.09$ & $-0.16\pm0.17$ & $-0.20\pm0.37 $&$-0.28\pm0.71$ & $-0.05$\\
    $g$  & $0.00\pm0.03$ & $0.00\pm0.05$ & $0.00\pm0.10$ & $0.00\pm0.18$&$0.00\pm0.37$&$0.01\pm0.74$ & 0.04\\
    $V$  & $0.05\pm0.03$ & $0.09\pm0.05$& $0.14\pm0.10$ & $0.21\pm0.18$&$0.26\pm0.36$&$0.18\pm0.71$ & 0.18\\
    $r$  & $0.22\pm0.05$ & $0.55\pm0.08$& $0.82\pm0.12$ & $1.32\pm0.19$&$2.98\pm0.39$&$3.90\pm0.76$ & 2.07\\
    $i$  & $0.43\pm0.06$ & $0.92\pm0.08$ & $1.47\pm0.13$ & $2.30\pm0.20$&$4.39\pm0.38$&$5.96\pm0.74$ & 1.68 \\
    $z_{\rm s}$  & $3.43\pm0.10$ & $1.60^{+0.15}_{-0.24}$ & $2.54\pm0.16$ & $3.89\pm0.26$ &$6.47\pm0.43$&$8.39\pm0.76$ & 2.81 \\
    \hline
    \end{tabular}

\end{table*}

\subsection{Timescale dependent lags}
\label{sec:timescale_lags}

Previous studies have noted that interband delays in AGN light curves can depend on the timescale of the variations. These timescale-dependent lags are typically investigated by de-trending the AGN light curves to separate faster and slower variations \citep[e.g.,][]{McHardy2018, Edelson19, JVHS_F9_2020,Cackett2023_Mrk817,Secunda2023,Edelson2024}. A more complete analysis examines the dependence of lags on Fourier frequency \citep{Panagiotou2025}. By characterising how the lag spectrum depends on the timescale of variations, we may recover new information on the response of the reverberating region to the initial impulse \citep[e.g.,][]{Starkey2016,Cackett2022_freq_rs}. Longer lags in the response to slower variations can be interpreted as arising from regions larger than the accretion disc, e.g., BLR \citep{KoristaGoad2001,Netzer2020,Netzer22, Lewin2023,Lewin2024} and/or above the disc plane, e.g., a wind \citep{Hagen2024}. 
  
For NGC\,4051, we investigate how the observed amplitudes and time delays depend on the timescale of the variations.
Specifically, we use the ROA timescale $\Delta$ to exclude faster variations from the light curve shape $X(t)$, and investigate how the rms amplitude $\Delta F(\lambda)$ and lag $\tau(\lambda)$ of the response at each wavelength vary with $\Delta$. 
\footnote{This is similar to the frequency-resolved lags used in the literature \citep{Lira2015,Cackett2022_freq_rs,Edelson2024}, which analyses the cross-spectrum between two light curves to measure delays versus temporal frequency
\citep{Uttley2014}.}
In addition to the {\sc PyROA} fit with $\Delta=0.65$\,d (presented in Sec.~\ref{sec:pyroa}), which is the global best fit to the data, we
 perform {\sc PyROA} fits with fixed values of $\Delta = [1.5, 3.0, 5.0, 10.0, 20.0]$\,d.
The collective lag measurements from this analysis are presented in Table\,\ref{tab:timedelays-two_years_sim} and displayed in Fig.\,\ref{fig:deltatrends}. Overall, we observe a trend for lags increasing with timescale. This can be interpreted as evidence for an asymmetric delay distribution with positive skew, for which slower variations have longer lags than faster ones \citep{Starkey2016}.
The faster variations $\Delta<1\,$~d have shorter lags $\tau<1$\,d, and increase with wavelength as expected for a compact accretion disc within which temperature decreases with increasing wavelength. Interestingly, longer lags develop for $\Delta>5$\,d, particularly in the $u$-band, which samples the Balmer continuum, and the redder $r$, $i$, and $z$ bands, which sample the Paschen continuum.  The $z$-band lag reaches a lag of 8\,days for $\Delta=20$\, d.
These trends with $\Delta$ and $\lambda$ suggest a mixture model with a compact multi-temperature accretion disc producing the prompt response from a region of order one light day in size, plus a more extended region of order 10 or more light days in size, responding as mostly free-bound and free-free hydrogen emission \citep{KoristaGoad2001,Korista2019,Netzer2020}. 

\subsection{{\sc MEMEcho} analysis}
\label{sec:memecho}

The 
{\sc PyROA} fits discussed above assume that each wavelength band $\lambda$ responds with a single time delay, $\tau(\lambda)$, so that the light curve shape $X(t)$ is the same at all wavelengths. 
The high quality of the LCO light curves warrants a more detailed analysis of the variations.
{\sc MEMEcho} 
allows each band to respond with a distribution of time delays, $\Psi(\tau|\lambda)$, while also having slowly-varying background flux $\bar{F}(\lambda, t)$.

\begin{figure*}
\centering
\includegraphics[width=\linewidth, trim={15mm 5mm 35mm 10mm}, clip]{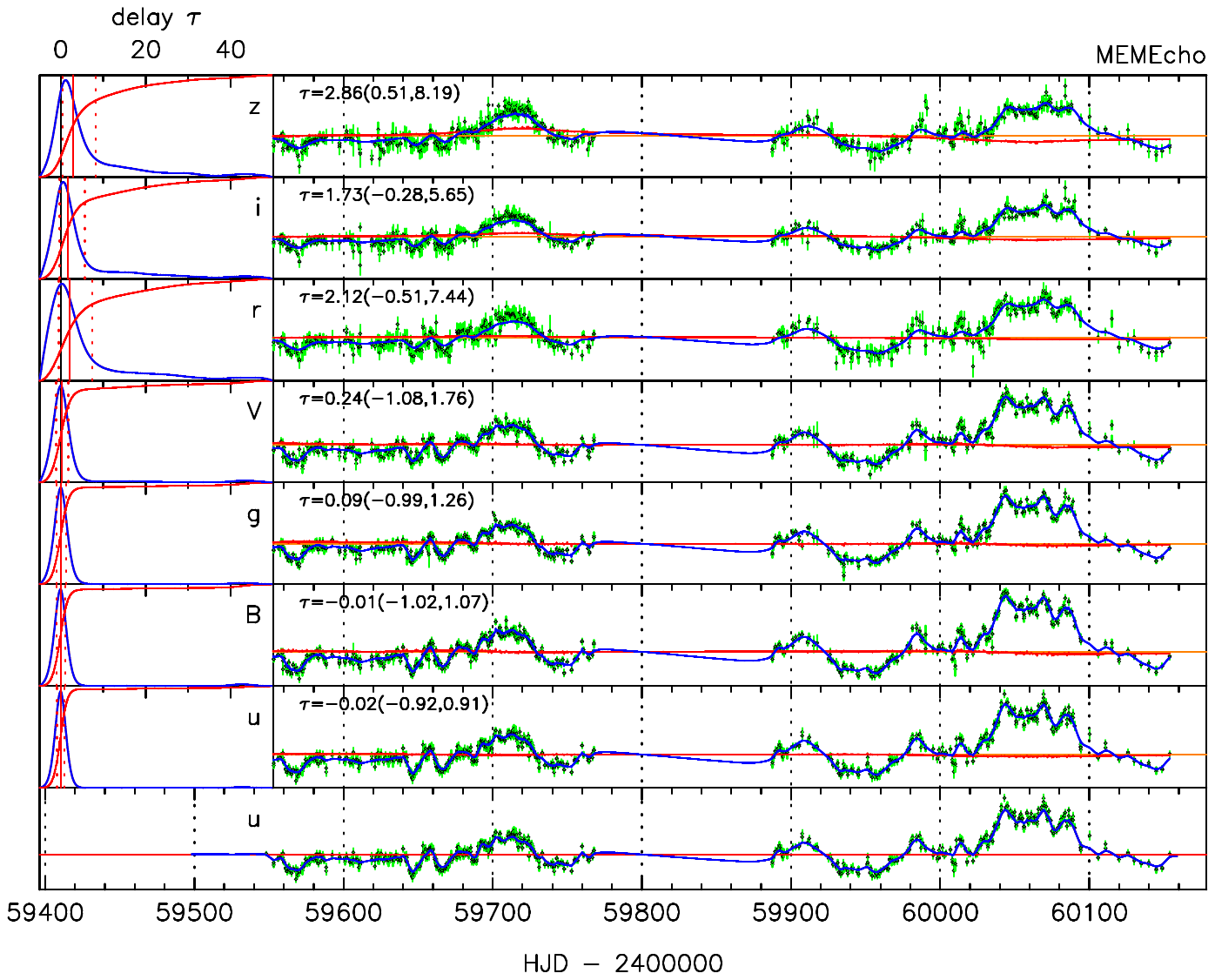}
\caption{\label{fig:Memecho}MEMecho fit to 7-band LCO light curves of NGC\,4051.
The observed fluxes (green points with error bars) are fitted with a linearised echo model with slow background variations, Eq.\,(\ref{eq:memecho}).
The driving light curve (bottom panel)
is convolved with 7 delay maps (left panels) to produce 7 echo light curves (right panels) plus
slowly-varying backgrounds (red curves).
MEMEcho adjusts the predicted light curves and delay maps (blue curves) to achieve a good fit ($\chi^2/N=1)$ 
to each of the light curves,
while maximising the entropy
to keep model curves positive and as smooth as possible. 
The delay map panels show the cumulative delay distributions (rising red curves). 
The median and quartiles of each delay map
are marked by vertical red solid and dashed lines.
These are listed both on the plot and
in Table\,\ref{tab:timedelays-two_years_sim}.
The delay maps all have a prompt component (disc) and the redder bands ($r$, $i$, $z_s$) have an extended response tail (BLR)
and slow variations (torus). We interpret this as due to the BLR and the obscuring torus, respectively. 
}
\end{figure*}

Our {\sc MEMEcho} fit to the LCO light curves is shown in
Fig.\,\ref{fig:Memecho}.
{\sc MEMEcho} 
models the driving light curve, in this case the $u$ band,
as a reference level plus variations,
$X(t) = \bar{X} + \Delta X(t)$.
We set $\bar{X}$ to the median of the $u$-band fluxes.
The echo light curves, one per wavelength band $\lambda$,
are modelled by the
{\it linearised} echo model:
\begin{equation}
\label{eq:memecho}
F(t,\lambda) = \bar{F}(\lambda,t) +
\int_{\tau_{\rm min}}^{\tau_{\rm max}}
\Psi( \tau | \lambda) \, \Delta X( t - \tau )\, d\tau
\ .
\end{equation}
The model parameters include
the driving light curve $X(t)$, the background light curves $\bar{F}(\lambda,t)$, and the
delay maps $\Psi(\tau|\lambda)$, all sampled at $0.1$\,d intervals.
The maximum entropy fit shown in Fig.\,\ref{fig:Memecho} adjusts these functions to achieve a good fit $\chi^2/N=1$ for the light curves in all bands.
The maximum entropy regularisation keeps these functions positive and smooth. In particular, the entropy steers each pixel value toward a default value that is a Gaussian-weighted average of neighbouring pixels.
The Gaussian widths that control the relative flexibility of the model functions were set to
0.2\,d for the delay maps, 1\,d for the driving light curve, and 10\,d for the background light curve.

The {\sc MEMEcho} fit results in Fig.\,\ref{fig:Memecho} show several notable features.
The delay maps (left panels) for all bands are dominated by a prompt response peak near zero delay.
The median delay, indicated by the vertical red lines in the left panels, increases with wavelength, reaching 2.8\,d in the $z_s$ band.
A long response tail, extending to $\sim30\,d$ or more, is present in the redder bands ($r$, $i$, and $z_s$).  
These bands also have small amplitude slow background variations (red curves) with a broad maximum near the end of Year\,1 and a broad minimum in the second half of Year\,2, thus a roughly 1-year timescale.
The fit to the bluer bands ($u$, $B$, $g$, $V$) was achieved with little or no extended response tail and with little or no variable background. 

We provisionally interpret the prompt response as reprocessing on the surface of the relatively compact accretion disc, the extended response tail as reprocessing in the BLR, and the slow variations as reprocessing on the inner edge of the dusty torus.

\begin{table*}
\caption{\label{tab:flux_flux_2years_delta0.65}Flux-flux analysis for NGC\,4051, 2 years simultaneously fitted with the optimised Gaussian memory function with a width $\Delta = 1.5$ d as measured by PyROA. The pivot wavelengths for each filter are in the observed frame. We present the de-reddened values for SED only by Galactic extinction \citep{SchlaflyFinkbeiner2011} with an $E(B-V)=0.011$.
    }

    \begin{tabular}{cccccccc}
    \hline
Filter & $\lambda$ & Galaxy flux & AGN$_\mathrm{bright}$ & AGN$_\mathrm{faint}$ & Galaxy flux & AGN$_\mathrm{bright}$  & AGN$_\mathrm{faint}$\\
&&&&&de-reddened&de-reddened&de-reddened\\
& (\AA) & (mJy) & (mJy) & (mJy) & (mJy) & (mJy) & (mJy) \\
    \hline 
    $u$ & 3580 & $0.037 \pm 0.037$ & $6.530 \pm 0.013$ & $3.698 \pm 0.010$ & $0.036 \pm 0.039$ & $6.856 \pm 0.014$ & $3.883 \pm 0.010$ \\
    $B$ & 4392 & $3.353 \pm 0.034$ & $5.975 \pm 0.037$ & $3.384 \pm 0.029$ & $3.494 \pm 0.035$ & $6.226 \pm 0.039$ & $3.526 \pm 0.030$ \\
    $g$ & 4770 & $5.402 \pm 0.034$ & $6.056 \pm 0.039$ & $3.343 \pm 0.030$ & $5.609 \pm 0.035$ & $6.288 \pm 0.041$ & $3.561 \pm 0.031$ \\
    $V$ & 5468 & $8.782 \pm 0.038$ & $6.811 \pm 0.047$ & $3.857 \pm 0.035$ & $9.059 \pm 0.039$ & $7.025 \pm 0.048$ & $3.978 \pm 0.036$ \\
    $r$ & 6215 & $13.314 \pm 0.045$ & $7.954 \pm 0.081$ & $4.505 \pm 0.052$ & $13.663 \pm 0.046$ & $8.163 \pm 0.083$ & $4.623 \pm 0.054$ \\
    $i$ & 7545 & $17.451 \pm 0.049$ & $8.711 \pm 0.089$ & $4.934 \pm 0.058$ & $17.789 \pm 0.050$ & $8.880 \pm 0.090$ & $5.029 \pm 0.059$ \\
    $z_s$ & 8700 & $23.138 \pm 0.053$ & $9.441 \pm 0.134$ & $5.347 \pm 0.089$ & $23.489 \pm 0.054$ & $9.584 \pm 0.136$ & $5.429 \pm 0.091$ \\  
\hline
\end{tabular}
\end{table*}

\section{Spectral energy distribution } \label{Sec:SED_flux}
\begin{figure*}
\centering
    \includegraphics[width = 0.49 \textwidth]{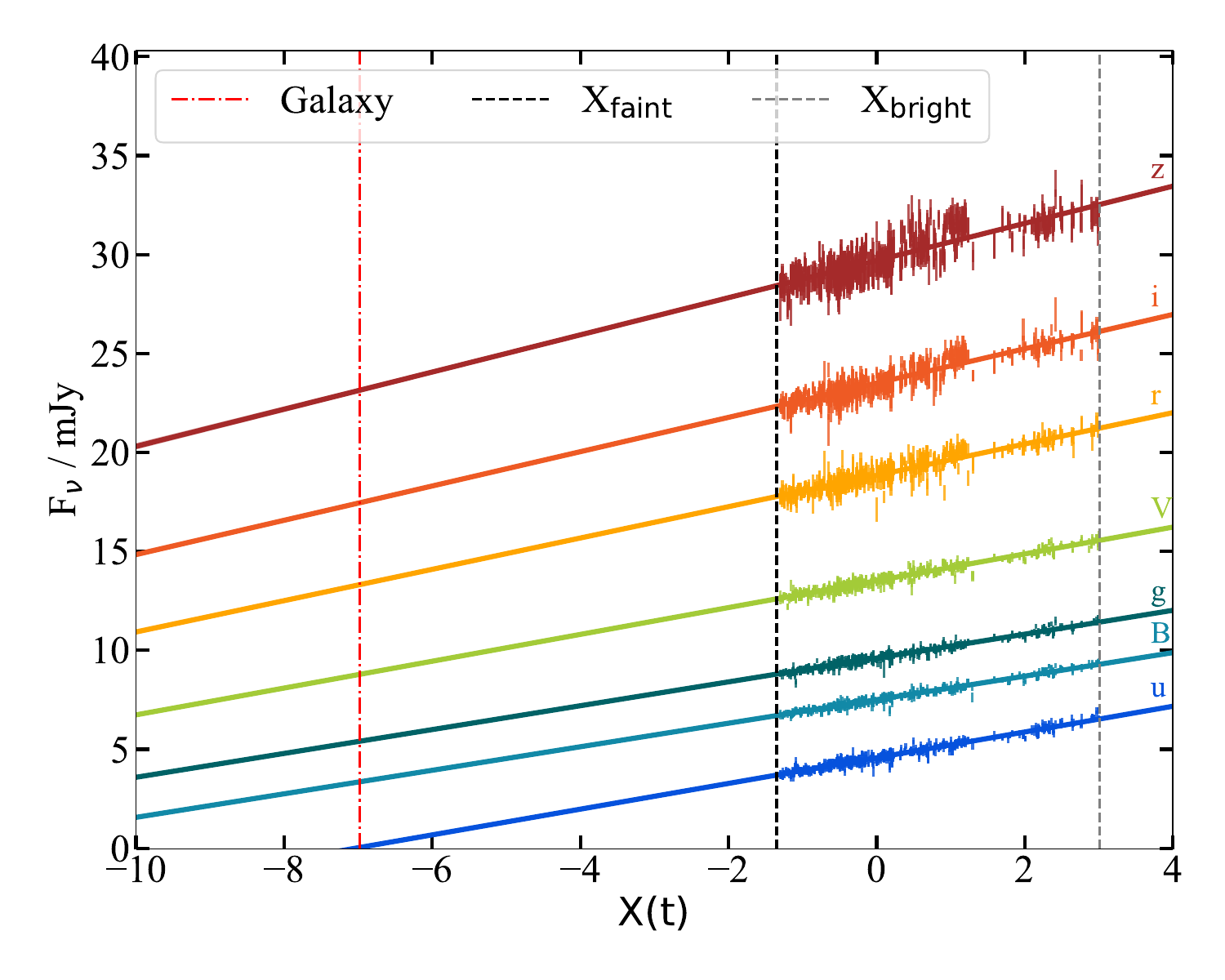}
    \includegraphics[width = 0.49 \textwidth]{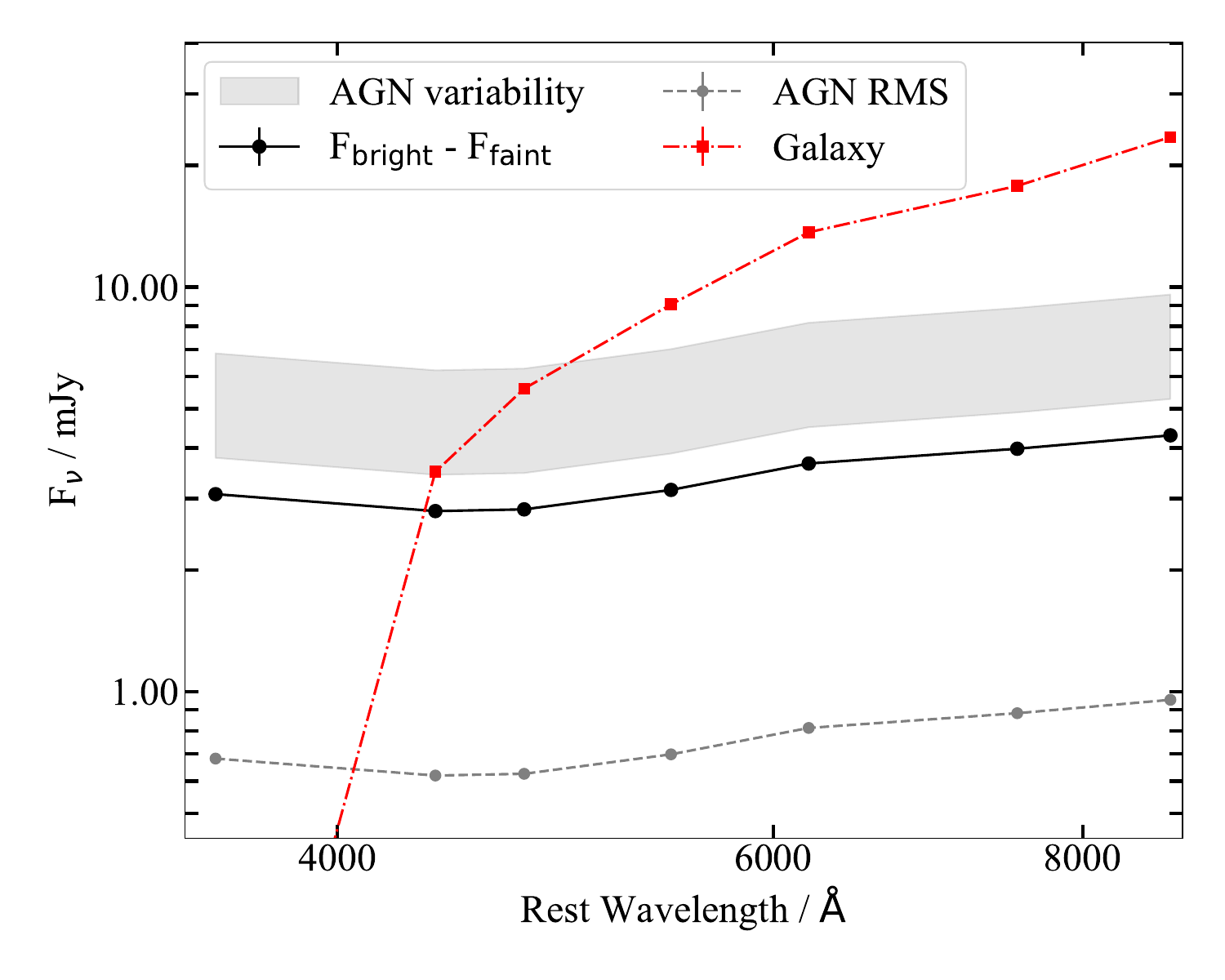}
    \caption{     \label{fig:flux_flux_SED_2sim_yr}
Results of the flux-flux analysis. The colours have the same meaning as in Fig.\,\ref{fig:2year_fit}. The left panel shows the observed fluxes as a function of the dimensionless light curve shape $X(t)$ {(from {\sc PyROA} fit)}, which represents the relative variability of the driving continuum normalized to its mean level.
The linear models fitted to the flux data adequately describe the variations in all bands, with the slopes $dF_\nu/dX$ given the RMS of the light curve at each wavelength. Extrapolating to fainter levels, the W2 flux is 1$\sigma$ above 0 at $X_{\rm g}$ (red dashed line). Evaluating the model fluxes at $X_{\rm g}$ gives the constant (host galaxy) SED. Subtracting this leaves the variable (AGN disc) SED, ranging from $X_{\rm F}$ at the faintest to $X_{\rm B}$ at the brightest during the campaign. The right panel shows the SEDs obtained from the flux-flux plot, corrected for Galactic dust extinction and reddening; $E(B-V) =0.011$ \citep{SchlaflyFinkbeiner2011}. The variable disc SEDs, faint, bright, and RMS, are marked in black and grey. The host galaxy SED (red) is much redder. The host galaxy flux inferred from the flux–flux analysis is consistent with the HST measurement of \citet{Bentz2013} ($f_{\nu,5100\text{\AA}}
\approx 7.6$ mJy), providing an independent confirmation of the constant component.}
\end{figure*}

We conduct a flux-flux analysis, summarised in Fig.\,\ref{fig:flux_flux_SED_2sim_yr}
and Table\,\ref{tab:flux_flux_2years_delta0.65}, 
to separate the variable (accretion disc) and constant (host galaxy) components of the AGN light
and thus to determine the SED of both emission components.
Our method is similar to that of \cite{Winkler1997}, except that we fit all bands simultaneously rather than in pairs.
We use the normalized light curve shape $X(t)$ and fit the light curves using:
\begin{equation}
    F(\lambda,t) = C(\lambda)+S(\lambda)\,X(t)\ ,
\end{equation}
where the flux at time $t$ is a linear combination of $C(\lambda)$ and $S(\lambda)$. 
{This linear {\sc PyROA}} model (left panel of Fig.\,\ref{fig:flux_flux_SED_2sim_yr})
is a good empirical description of the correlated variations.
The slope $S(\lambda)$ at each wavelength gives the rms SED of the variable (AGN) component.
We evaluate the model 
at $X_{\rm faint}$ and $X_{\rm bright}$ to estimate SEDs at the faintest and brightest states sampled by the monitoring campaign,
and we use the linear model to
extrapolate to fainter or brighter states.
The SED of the constant (host galaxy) contribution is thus estimated by
\begin{equation}
F_{\rm gal}(\lambda) = C(\lambda) + S(\lambda) \, X_{\rm gal}\ ,
\end{equation}
where (by convention) $X_{\rm gal}$ is set where the extrapolated model for the first band is $1\sigma$ above zero.

Table\,\ref{tab:flux_flux_2years_delta0.65} reports the resulting SEDs, before and after our correction for Milky Way dust, for the galaxy and the AGN disc components in the bright and faint states.
The right panel of
Fig.\,\ref{fig:flux_flux_SED_2sim_yr} summarizes the SED results. The host galaxy component is marked in red, and a light gray envelope indicates the range of variations in the AGN disc component.
The relatively red galaxy SED is evident. 

Fig.\,\ref{fig:deltatrends} (bottom panel) shows the faint and bright AGN disc SEDs determined from the mean and rms of light curves measured from {\sc PyROA} fits with different smoothing widths $\Delta$. 
The spectral shape is independent of $\Delta$ and suggests free-bound emission with the Balmer continuum sampled by the $u$-band and the Paschen continuum by the redder bands, although other interpretations are possible.
The absolute scaling of the inferred bright disc SED shape is largely independent of $\Delta$. 
The faint disc SED is also independent of $\Delta$, except
that the flux is fainter for the lowest and brighter for the highest values considered (0.65\,d and 20\,d).
{As shown in the timescale-dependent lag analysis (Sec.\,\ref{sec:timescale_lags}),
the interband lags transition from short delays ($\tau < 1$\,d), which are consistent with a
compact accretion disc, to significantly longer delays for $\Delta \gtrsim 5$\,d,
indicating contributions from a more extended reprocessing region (Fig.\,\ref{fig:deltatrends}). We therefore adopt $\Delta = 10$\,d to model the slow variations, as this timescale lies well above the $\Delta \sim 5$\,d transition where the longer lags emerge, effectively isolating the response of the extended region while filtering out the faster, disc-dominated variability. The {\sc PyROA} fit for
$\Delta=10$\,d is shown in Fig.\,\ref{fig:2year_wslow}.}


\subsection{Extending the SED into the UV with archival \swift\ data}
\label{sec:swift}

Most of the \swift\ observations of NGC\,4051 are not contemporaneous with our intensive monitoring with LCO.
Nevertheless, they cover a wider range of luminosities and enable us to extend the flux-flux analysis into the UV energy range and thereby to better define the SED shape. The \swift\ UVOT observations at some epochs include the $B$ band, which is similar to its counterpart in LCO. 
We can use the solution obtained by our best fit in Sec.~\ref{sec:pyroa} to pair these $B$-band observations and map each \swift\ epoch directly to a particular $X(t)$ value. We then use this $X(t)$ for the other UVOT bands at the same epoch, enabling us to include the UVOT bands in the flux-flux analysis.

\begin{figure*}
\centering
    \includegraphics[width=0.45\linewidth]{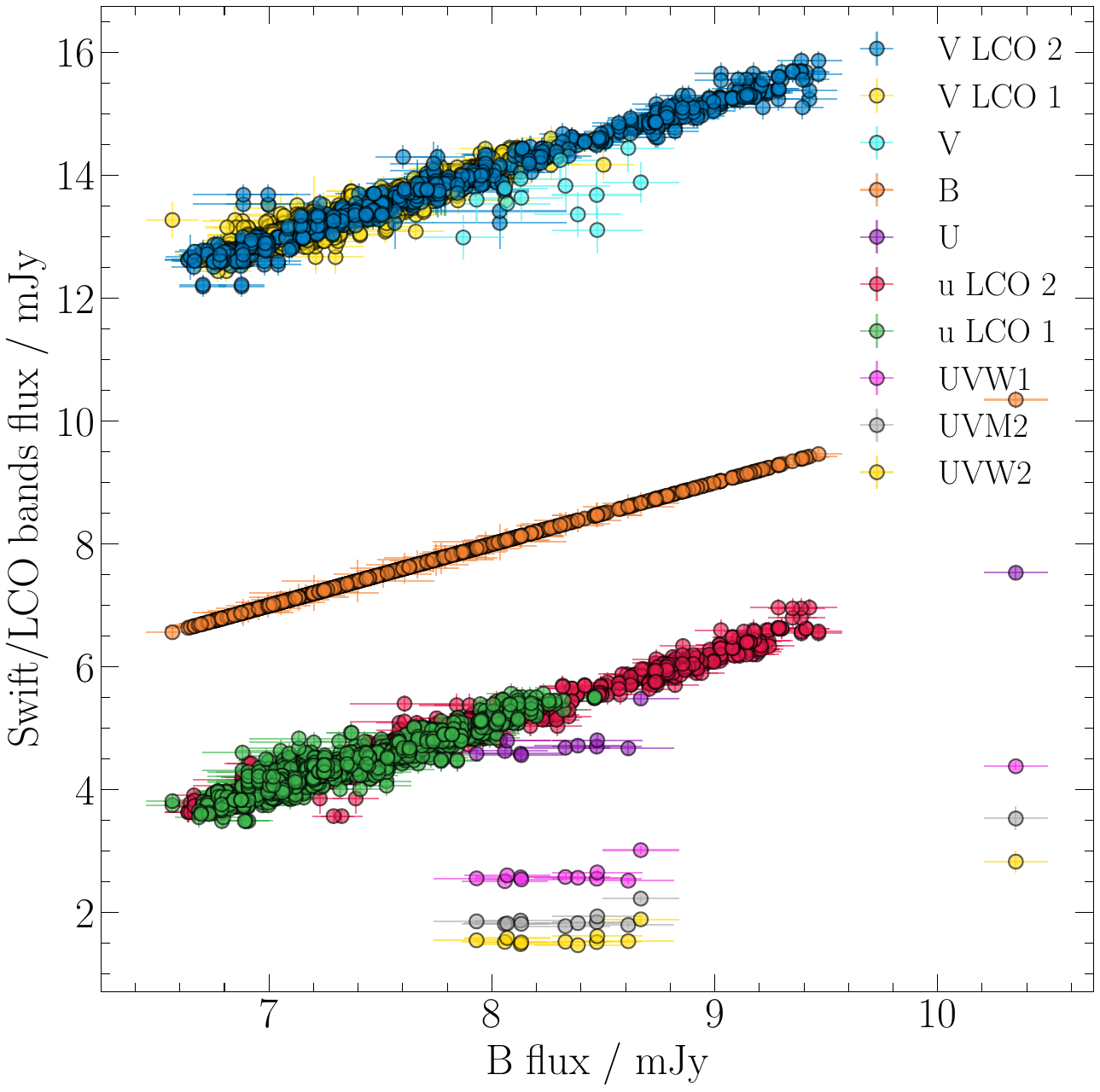}
    \includegraphics[width = 0.45\linewidth]{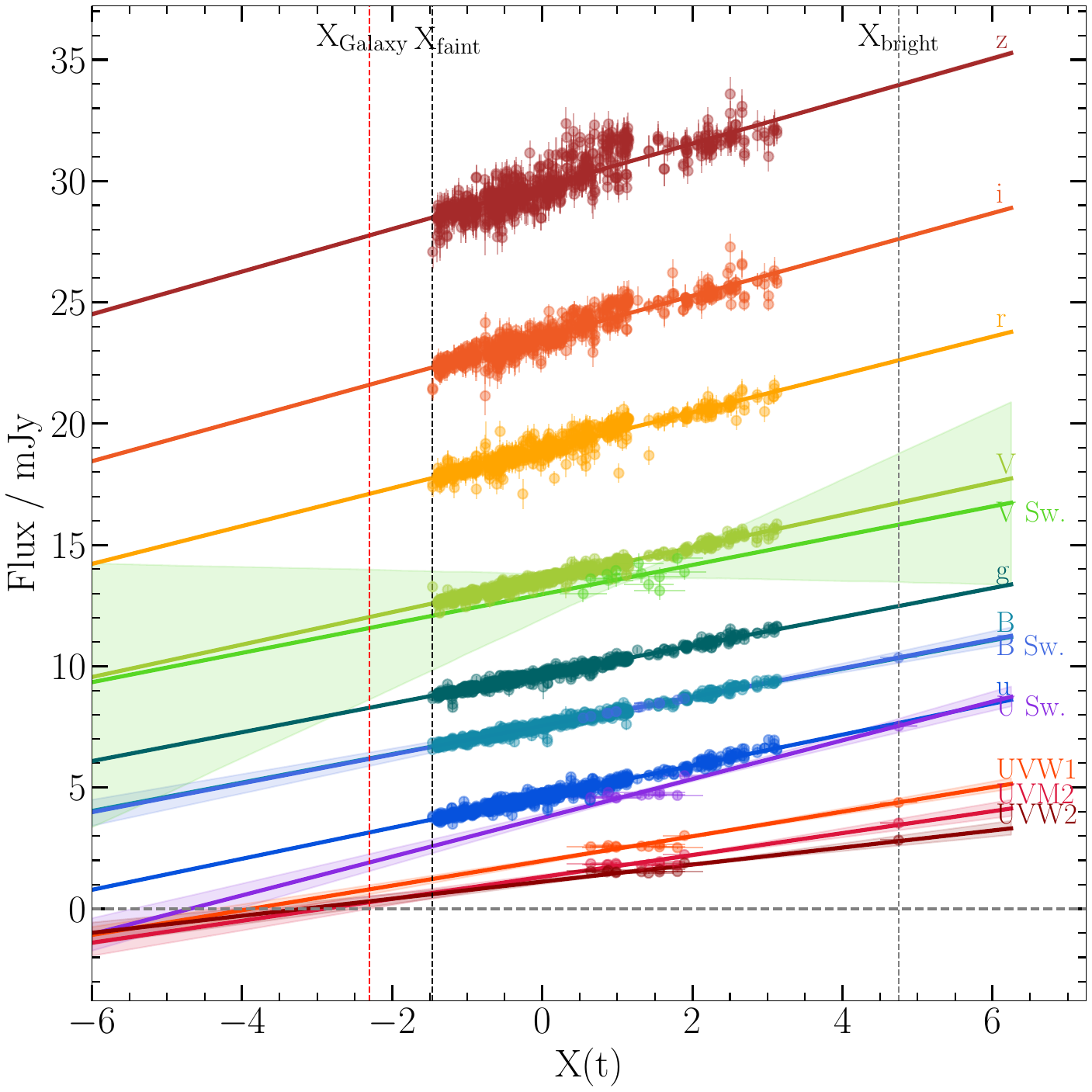}
    \caption{\label{fig:swift} Flux-flux analysis including archival \swift\ data (left). We use the $B$-band (orange) as a proxy and reference band to match $X(t)$. 
    Flux-flux analysis of simultaneous \swift\ and LCO data (right). The data are presented in a color-coded format analogous to that of Fig.\ref{fig:2year_fit}, with the incorporation of \swift\ bands, UVW2, UVM2, UVW1, $U$, $B$, $V$, dark red, crimson, orange red, blue violet, royal blue, lime green, respectively. Each data point is associated with a 68\% confidence interval. It is evident that the variable $V$-band is the sole band that has resulted in an augmentation of the uncertainty envelope, a consequence of the substantial dispersion of the data points.
        }
\end{figure*} 

We show this pairing of bands to the B-band and their linear relationship in the left panel of Fig.~\ref{fig:swift}. The linear relations defined by the LCO $u$ and $V$ fluxes vs LCO $B$ flux are evident. The archival \swift\ UVOT data, at the bottom of the diagram, are sparse in time sampling but do suggest that a linear relation may hold as the source brightens and fades. From this, we conclude that the archival \swift\ data brightness range overlaps with that of the LCO data, and extends it at one epoch to even brighter states
for all but the V band.
 We performed a linear fit using $\chi^2$ as our goodness-of-fit statistic, as shown in the right panel of Fig.~\ref{fig:swift}. 
As before, the best-fit slopes give the rms amplitude of the variations in each band, to thus define the variable AGN disc SED. 

Extrapolating to fainter levels, the linear trend fitted to the UVW2 fluxes revises our estimate of $X_{\rm gal}$ {(red dashed line), \mbox{$X_{\rm Galaxy}=-2.30$}} as indicated in the right panel of Fig.\,\ref{fig:swift}.
This analysis assumes that the linear flux-flux model holds for the \swift\ bands at earlier times and at lower and higher brightness levels than observed in the intensive monitoring. The linear flux-flux model has been shown to be a good representation of UV and optical variations observed in other AGN \citep[e.g.,][]{Fausnaugh2016RM_NGC5548,Cackett20,JVHS_F9_2020}.
However, the linear extrapolation to low levels could break down for the bluest bands because colour changes are expected as the peak temperature of the disc decreases. 
The data should then show concave-up curvature deviating from the extrapolated linear model.
Our revised $X_{\rm gal}$ could thus be set too high.

\begin{figure}
\centering
\includegraphics[width=0.45\textwidth,trim= 5mm 50mm 0mm 0mm,clip]{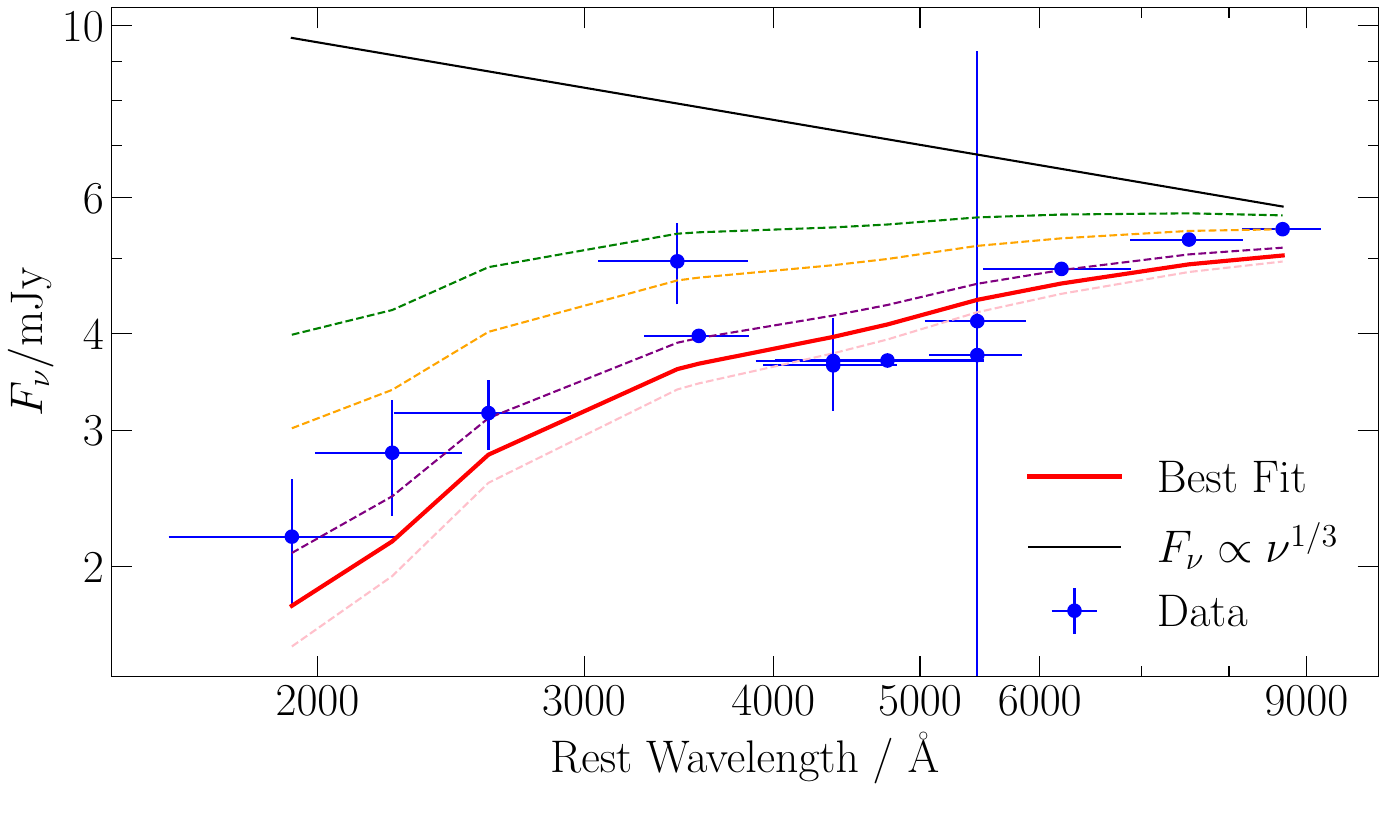} \includegraphics[width=0.5\textwidth,trim=16mm 10mm 40mm 30mm,clip]
{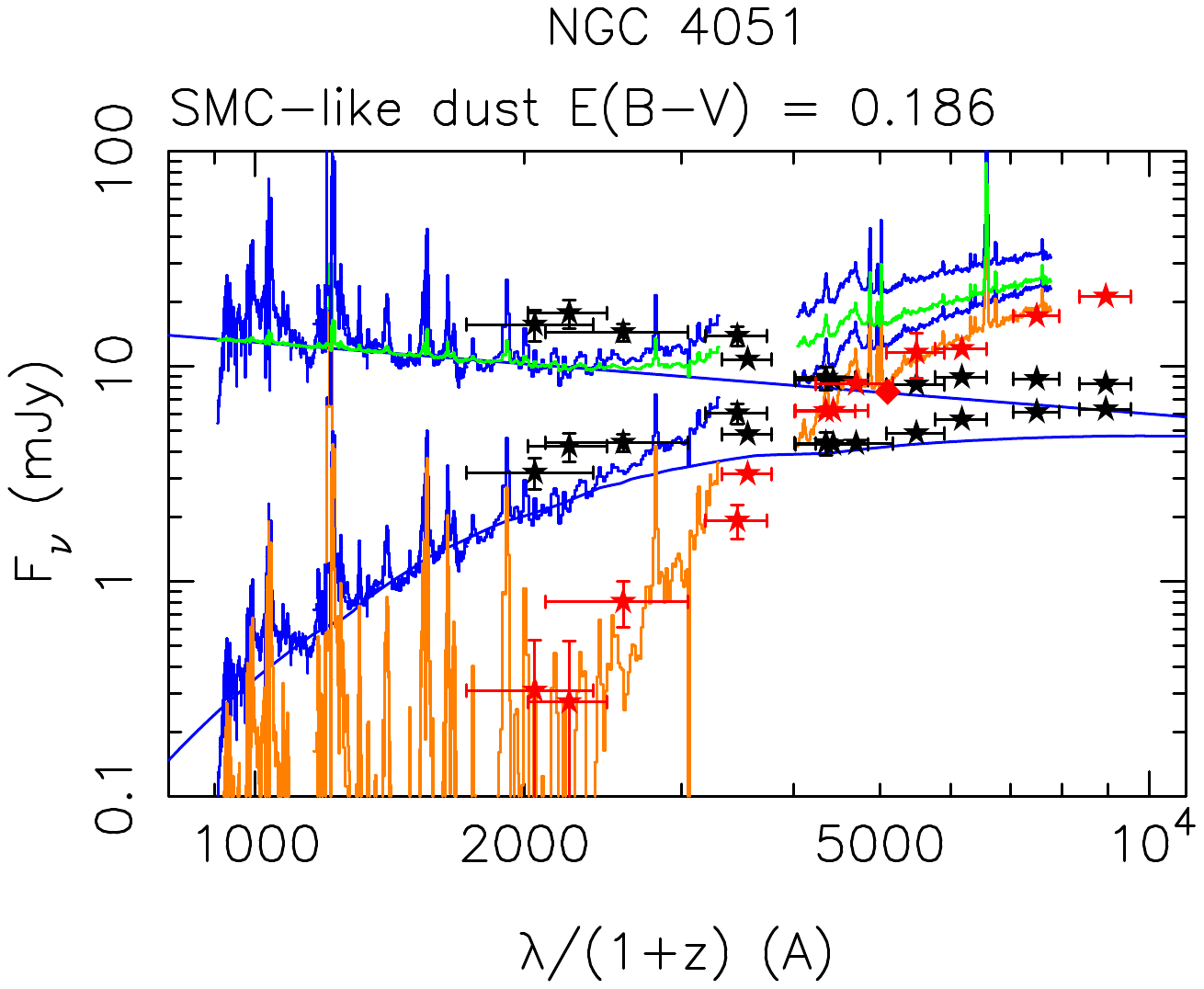}
\caption{\label{fig:SED_Swift_LCO_b_f}
{\it Top:} The joint LCO and \swift\ SED of NGC\,4051 from our flux-flux analysis (Table\,\ref{tab:fluxflux_LCO+Swift}),
corrected for line-of-sight Milky Way extinction of $E(B-V)=0.011$,
and compared with a power-law accretion disc model, $F_{\nu} \propto \nu^{1/3}$, 
after applying additional SMC-like dust extinction. 
We test a fine grid of $E(B-V)$ values
and show only a few, 0.10, 0.13, 0.17, 0.20, green, orange, purple, and pink, respectively. 
The best fit (red) is $0.186\pm 0.025$. 
\\ {\it Bottom:} The SED of NGC\,4051 (blue histograms) assembled
from FUSE, HST/STIS, IUE, and optical spectra
\citep{Kaspi2004},
before and after correcting for SMC-like dust extinction.  
The accretion disc power-law, $F_\nu\propto\nu^{1/3}$, fits the continuum.
Subtracting the extincted power-law SED leaves an estimate for the host galaxy SED (orange).
This agrees well with the galaxy SED 
(red stars with error bars) from our \swift+LCO flux-flux analysis (Table\,\ref{tab:fluxflux_LCO+Swift}).
The bright AGN fluxes (black stars with error bars) are $\sim0.2$\,dex above the power-law disc SED, with the same spectral slope.
}
\end{figure}


Table\,\ref{tab:fluxflux_LCO+Swift}
gives the updated SEDs including both LCO and \swift\ data, before and after 
our Milky Way dust correction.
Fig.~\ref{fig:SED_Swift_LCO_b_f}
presents the resulting mean SED of the variable AGN disc, with the constant (host galaxy) SED subtracted. 
This is significantly redder than the $F_\nu\propto\nu^{1/3}$ power law expected for a standard thin blackbody accretion disc.
Comparing the AGN disc SED from the flux-flux analysis of LCO data only (right panel of Fig.\,\ref{fig:flux_flux_SED_2sim_yr}) with that from the joint analysis of \swift\ and LCO data (Fig.\,\ref{fig:SED_Swift_LCO_b_f}), the disc SED shape and flux level are very similar for the LCO bands. This is expected given the much smaller uncertainties on the LCO fluxes. Despite larger uncertainties for the \swift\ bands, the $u$-band excess flux and the relatively red SED are more clearly defined with the \swift\ band data included. 

\begin{table*}
\caption{\label{tab:fluxflux_LCO+Swift}Combined LCO and \swift\ flux-flux analysis for NGC\,4051. The galaxy and AGN fluxes have been de-reddened by Galactic absorption with an $E(B-V) = 0.011$ \citep{SchlaflyFinkbeiner2011}.}
    \begin{tabular}{cccccccc}
    \hline
Filter & $\lambda_{\rm obs}$ & Galaxy flux & AGN$_\mathrm{bright}$ & AGN$_\mathrm{faint}$ & Galaxy flux  & AGN$_\mathrm{bright}$ & AGN$_\mathrm{faint}$ \\
&&&&&de-reddened &de-reddened &de-reddened \\
    & (\AA) & (mJy) & (mJy) & (mJy) & (mJy) & (mJy) & (mJy) \\
    \hline
    UVW2 & 1928 & $0.310 \pm 0.221$ & $2.478 \pm 0.409$ & $0.294 \pm 0.049$ & $0.400 \pm 0.285$ & $3.196 \pm 0.528$ & $0.380 \pm 0.063$ \\
    UVM2 & 2246 & $0.274 \pm 0.252$ & $3.182 \pm 0.476$ & $0.378 \pm 0.057$ & $0.365 \pm 0.336$ & $4.245 \pm 0.635$ & $0.504 \pm 0.076$ \\
    UVW1 & 2600 & $0.802 \pm 0.192$ & $3.582 \pm 0.324$ & $0.426 \pm 0.038$ & $0.987 \pm 0.236$ & $4.405 \pm 0.398$ & $0.524 \pm 0.047$ \\
    U & 3465 & $1.914 \pm 0.348$ & $5.629 \pm 0.594$ & $0.669 \pm 0.071$ & $2.052 \pm 0.373$ & $6.036 \pm 0.637$ & $0.717 \pm 0.076$ \\
    B & 4392 & $6.178 \pm 0.286$ & $4.182 \pm 0.503$ & $0.497 \pm 0.060$ & $6.480 \pm 0.300$ & $4.386 \pm 0.528$ & $0.521 \pm 0.063$ \\
    V & 5468 & $11.560 \pm 2.730$ & $4.257 \pm 5.489$ & $0.506 \pm 0.652$ & $11.954 \pm 2.823$ & $4.402 \pm 5.676$ & $0.523 \pm 0.675$ \\
    \hline
    u & 3580 & $3.151 \pm 0.010$ & $4.508 \pm 0.027$ & $0.536 \pm 0.003$ & $3.367 \pm 0.011$ & $4.817 \pm 0.029$ & $0.573 \pm 0.003$ \\
    B & 4392 & $6.202 \pm 0.009$ & $4.129 \pm 0.022$ & $0.491 \pm 0.003$ & $6.506 \pm 0.009$ & $4.331 \pm 0.023$ & $0.515 \pm 0.003$ \\
    g & 4770 & $8.290 \pm 0.009$ & $4.189 \pm 0.025$ & $0.498 \pm 0.003$ & $8.643 \pm 0.010$ & $4.368 \pm 0.026$ & $0.519 \pm 0.003$ \\
    V & 5468 & $12.028 \pm 0.012$ & $4.710 \pm 0.030$ & $0.560 \pm 0.004$ & $12.438 \pm 0.012$ & $4.870 \pm 0.031$ & $0.579 \pm 0.004$ \\
    r & 6215 & $17.108 \pm 0.021$ & $5.503 \pm 0.055$ & $0.654 \pm 0.007$ & $17.581 \pm 0.021$ & $5.655 \pm 0.057$ & $0.672 \pm 0.007$ \\
    i & 7545 & $21.605 \pm 0.022$ & $6.006 \pm 0.058$ & $0.714 \pm 0.007$ & $22.041 \pm 0.023$ & $6.127 \pm 0.060$ & $0.728 \pm 0.007$ \\
    z$_{\rm s}$ & 8700 & $27.762 \pm 0.030$ & $6.194 \pm 0.088$ & $0.736 \pm 0.010$ & $28.206 \pm 0.030$ & $6.294 \pm 0.089$ & $0.748 \pm 0.011$ \\
    \hline
\end{tabular}

\end{table*}

\subsection{The X-ray SED}
\begin{figure*}
\centering
    \includegraphics[width=0.8\columnwidth]
    {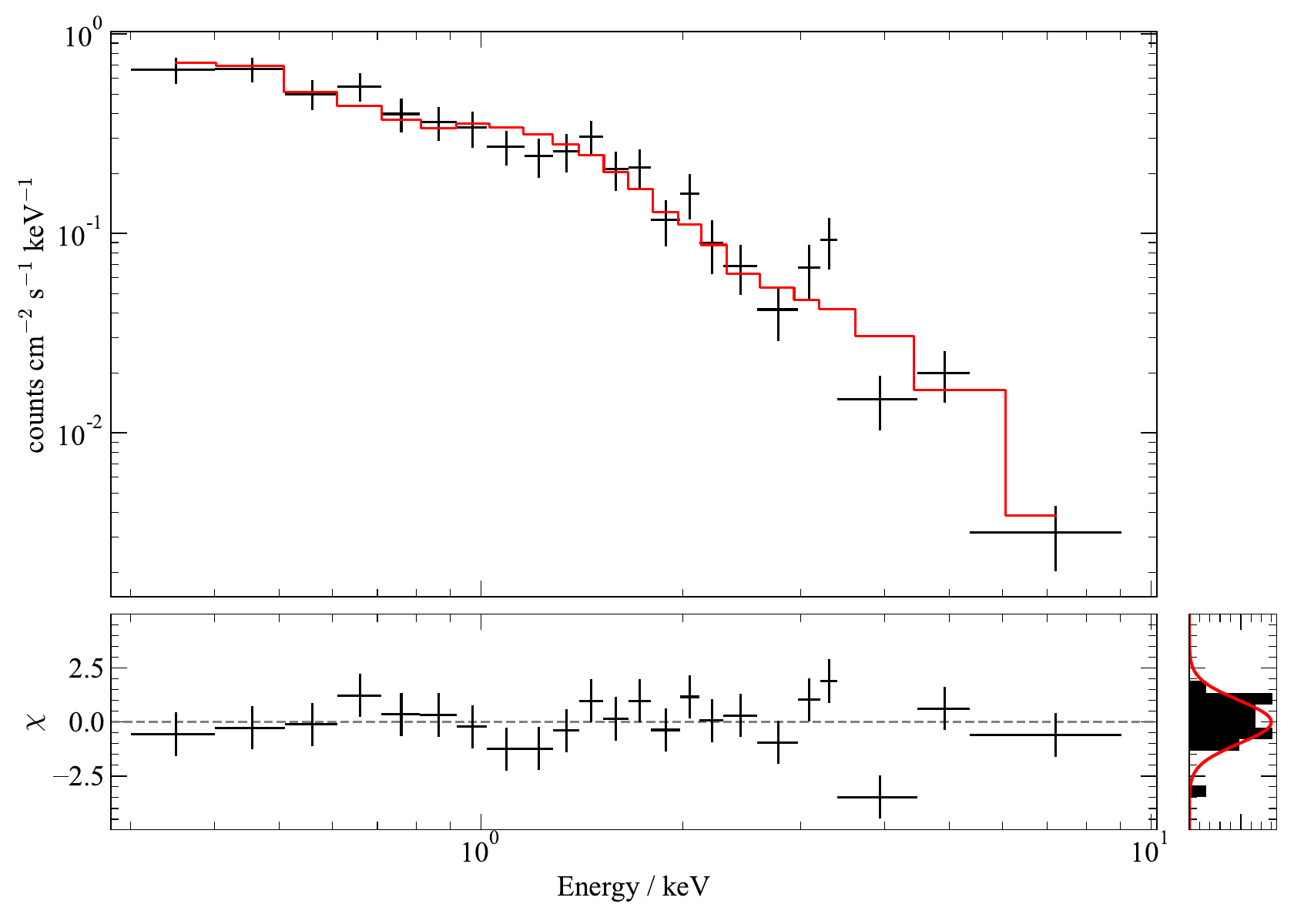}
\includegraphics[width=1.25\columnwidth]
    {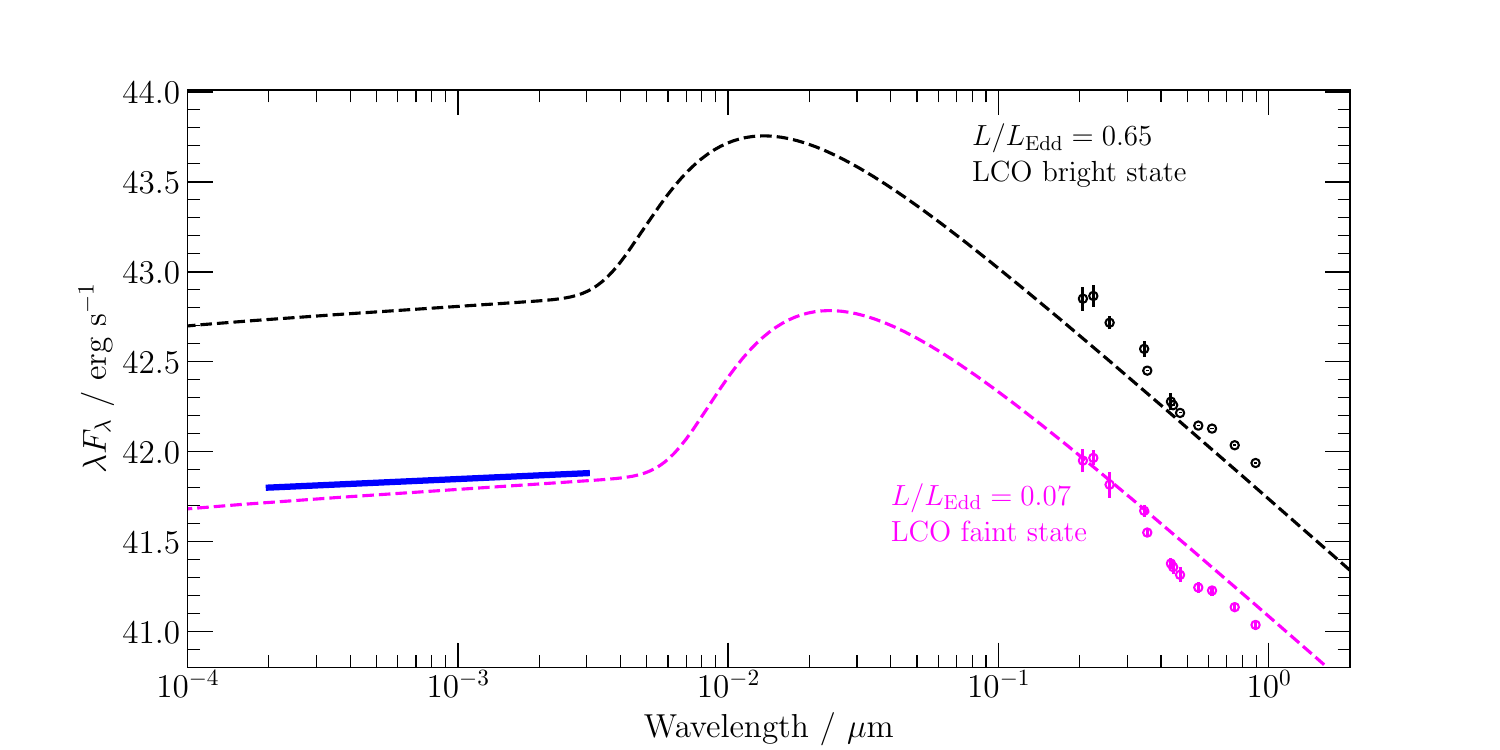}
    \caption{    \label{fig:Xray_spectrum}\textit{Left: }X-ray spectrum of NGC\,4051 and the best fitting model. Top: The absorbed power-law model is shown in red. The normalised residuals are shown in the bottom panels. The right-hand panel shows a histogram of the normalised residuals in comparison to a standard Gaussian distribution. \textit{Right:} The unfolded X-ray spectrum (blue) seen in context of the full SED. We show how the expected emission from a $L/L_{\rm Edd} \sim 0.07$ AGN \citep{Kubota2018} can reproduce both the X-ray emission and the faint state from the flux-flux analysis in Sec.~\ref{Sec:SED_flux} assuming SMC-like dust in the line-of-sight. }

\end{figure*}  
As mentioned in Sec.~\ref{sec:swift}, there were several \swift\ observations taken during our monitoring campaign. Therefore, to complement the UV to IR SED, we constructed an average X-ray spectrum with the \swift\ online tool \citep{Evans:2009}, shown in Fig.~\ref{fig:Xray_spectrum}. The X-ray compact source is absorbed by an ionised outflow, as revealed in previous detailed X-ray spectroscopic studies \citep{Nucita2010,Alston2013,Kumari2023}. Hence, we have analysed the X-ray spectrum with {\sc xspec} 12.11.1 \citep{Arnaud1996}, using a model that includes galactic absorption {\sc phabs} fixed at the Galactic line-of-sight column density $1.2\times10^{20}$ cm$^{-2}$ 
\citep{HI4PI2016}, a warm ionised partial-covering absorber {\sc zxipcf}, and a power-law component. Our full model within {\sc xspec} is {(\sc phabs * zxipcf * cflux * zpowerlw)}. The data are well described by the model with a best-fit statistic $C_{\rm stat}/{\rm dof} = 187.8/222$, as shown in Fig.~\ref{fig:Xray_spectrum}. The best-fit parameters and their 68\% confidence intervals are collected in Table~\ref{tab:xray_results}. Higher SNR observations have revealed the presence of a soft-excess \citep{Nucita2010,Alston2013}. However, due to the lower SNR of the \swift\ XRT data, adding an additional blackbody was not preferred by the data, pushing its normalisation to very low values. We find that the X-ray luminosity during our campaign was lower by a factor of $\sim2$ than the state observed in \citet{Nucita2010} where $L_{\rm 2\, keV}=(2.91\pm0.01)\times 10^{-29}$ \ergs cm$^{-2}$Hz$^{-1}$. It is unclear if the lack of a soft excess is due to the change in the overall luminosity of the AGN or an effect of the signal-to-noise. Recently, \cite{Serafinelli2025ApJ} found that in their faintest \textit{XMM-Newton} observations of NGC\,4051, the soft-excess was also not required.

The right panel of Fig.~\ref{fig:Xray_spectrum} includes the X-ray spectrum model in the wide perspective SED. It is clear that the X-rays are fainter than the UV/optical emission from the disc (albeit by a factor of a few). Therefore, large changes in the X-ray emission would be needed to drive the observed UV/optical/NIR variability. Thus, the origin of the bulk driving luminosity must come from the un-observed extreme-UV \citep{Gardner17,Kubota2018,Hagen2024}. The inferred Eddington ratio found in Sec.~\ref{sec:bowlmodel} suggests a $L/L_{\rm Edd}\sim4$\%, close to the transition limit seen in AGN population studies where the soft excess disappears \citep{Hagen2024b}. Further deep X-ray observations in this state should reveal (or confirm the lack of) the soft-excess contribution in this AGN.

{
\renewcommand{\arraystretch}{1.2}
\begin{table}
    \centering
\caption{\label{tab:xray_results}Best fit parameters of the X-ray spectrum of NGC\,4051.}
    \begin{tabular}{c|c|c}
    \hline
    Parameter & units & values\\
    \hline
$N_{\rm H}$ & $10^{22}$ cm$^{-2}$ & $0.012^{\star}$\\
$N_{\rm H}$ & $10^{22}$  cm$^{-2}$ & ${4.2}^{+16.5}_{-3.2}$\\
Cover Fraction &  & $0.54^{+0.22}_{-0.30}$\\
$\log(\chi)$ &  & ${1.3}^{+1.9}_{-2.9}$\\
$\Gamma$ &  & $2.15^{+0.28}_{-0.11}$\\
$\log(F_{\rm 0.5-10\, keV})$ & erg cm$^{-2}$ s$^{-1}$ & ${-10.13}^{+0.17}_{-0.10}$\\

C-Stat / dof &   & 187.8 / 222\\
\hline
         
    \end{tabular}\\
\begin{flushleft}{$^{\star}$   Fixed value for Galactic column density $N_{\rm H}$  
}
\end{flushleft}
    
\end{table}
}


\section{Discussion}
\label{Sec:Discussion}

The observed delay spectrum and AGN disc SED provide crucial tests for determining the size of emitting regions and the accretion flow geometry. 
These results are particularly valuable for testing competing accretion disc models and for inferring the temperature profile, and thus the mass accretion rate.

\subsection{Size of the reverberating region} \label{discsize}


A thin blackbody accretion disc with $T\propto r^{-3/4}$ corresponds to a delay spectrum with $\tau\propto \lambda^{4/3}$ \citep{Cackett2007reprocessing_model}.
Accordingly, we fit the delay spectrum with a power-law model:
\begin{equation}\label{eq:theory}
    \tau = \tau_0 \,\left[ \left(\frac{\lambda}{\lambda_0} \right)^\beta - {y_0}  \right]\ ,
\end{equation}
where $\tau_0\, c$ estimates the disc size at the reference wavelength $\lambda_0$, $\beta$ is the power-law index, and $y_0$ allows the model lag to depart from the observed lag at $\lambda_0$. We use the lag measurements for the combined two years of data, with $\Delta$ = 1.5\,d (see Table\,\ref{tab:timedelays-two_years_sim}). We optimise parameters using a Markov Chain Monte Carlo procedure as implemented by {\sc emcee} \footnote{https://emcee.readthedocs.io/en/stable/} with $10^4$ iterations, and estimate parameter uncertainties from the marginalised posterior distributions.

\begin{figure}
\centering
    \includegraphics[width = 0.49 \textwidth]{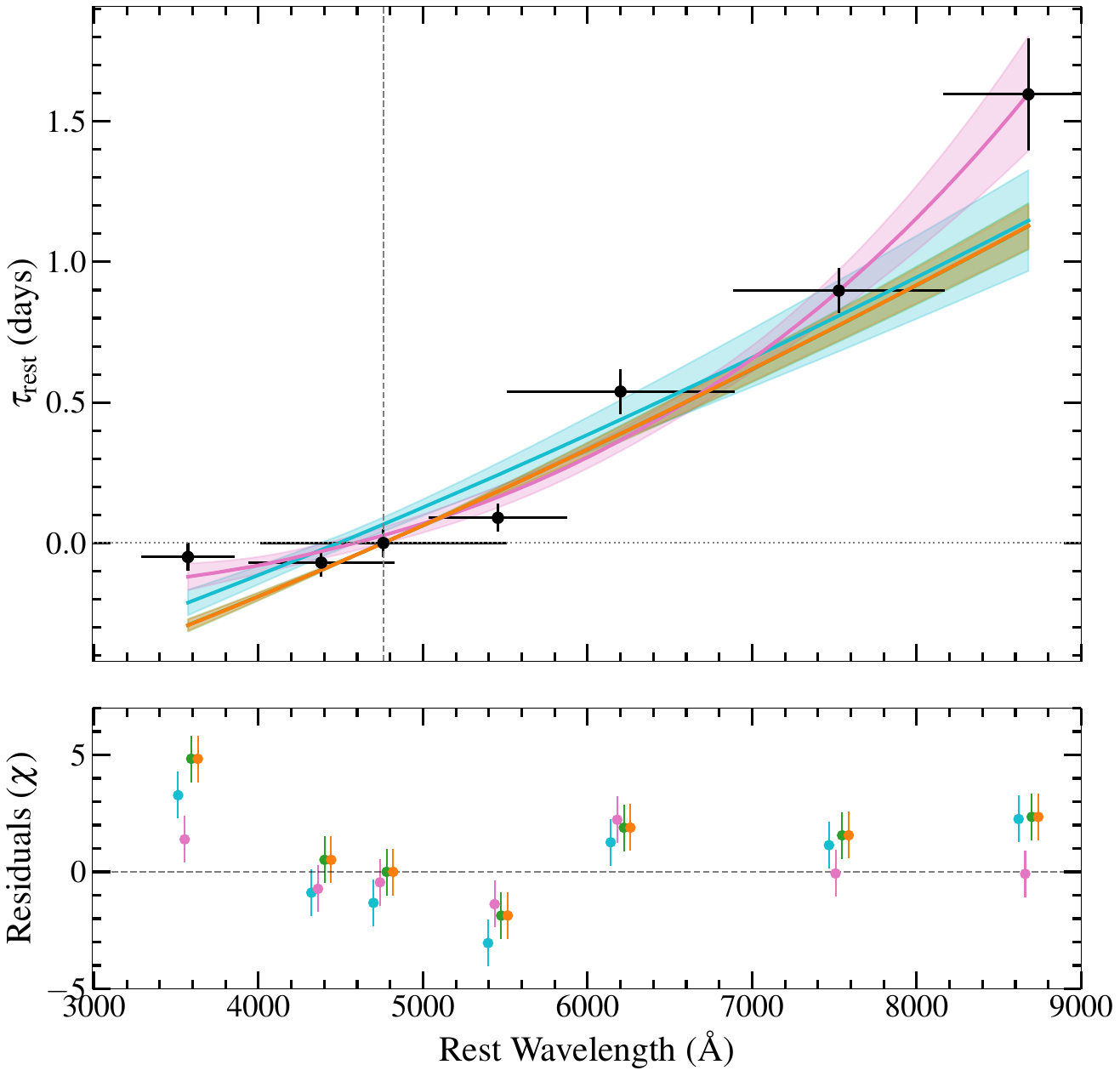}
\caption{\label{fig:delay_spectrum}Delay spectrum fit, after shifting to the rest frame, for the two years simultaneously as measured by {\sc PyROA}. The figure shows the best power-law fit with $\beta=4/3$ (cyan line) and its $1\sigma$ confidence envelope, as well as the fit where $\beta$ is left as a free parameter (magenta line). The best parameters for these fits are presented in Table\,\ref{tab:delay_spectrum_model}. Based on \citet{Kammoun2021} model, we check two cases: the best fit model with free parameters, when the colour correction was set free (green curve), we obtain $f_{\rm col}=2.4 \pm 0.3$, with accretion rate $\dot{m}_{\rm Edd} = 0.81\pm0.05$ for corona height; $h_{\rm c}=16^{+29}_{-12}$ $r_g$. The orange curve represents the delay spectrum when colour correction is fixed $f_{\rm col}$ = 1.7 with the best $\dot{m}_{\rm Edd} =0.92\pm 0.02$ and $h_{\rm c} =17^{+29}_{-12}$ $r_g$. The normalised residuals for each fit are shown in the lower panels. Please note that in regions where the fitted curves overlap closely, their colours blend together visually, making them difficult to distinguish. To improve clarity, a horizontal offset of 50 $\angstrom$ has been applied to the residuals.}
\end{figure} 

Initially, we fix $\beta=4/3$, as expected from the standard lamp-post irradiation model for a thin accretion disc \citep{Cackett2007reprocessing_model}, which predicts a lag-wavelength dependence of $\tau \propto \lambda^{4/3}$. The best fit is shown in Fig.~\ref{fig:delay_spectrum} and its parameter values are displayed in Table~\ref{tab:delay_spectrum_model}. We find $\tau_0 = 0.88 \pm 0.07$\,d, and $y_0 = 0.92 \pm 0.03$, with a best-fit statistic $\chi^2_\nu = 6.11$. This high $\chi^2_\nu$ is due largely to the $u$-band lag excess, as seen in the lower panel of Fig.\,\ref{fig:delay_spectrum}.

Several studies have fixed the power-law index at $\beta = 4/3$ to test consistency with the standard accretion disc model, with many finding good agreement \citep[e.g.,][]{Cackett18,Edelson19,JVHS_F9_2020,Vincentelli2021,Donnan23_PG1119} as well as the excess in particular bands, e.g., $u$-band.
For a standard thin blackbody accretion disc model, the predicted size
at wavelength $\lambda_0$
\citep{Fausnaugh2016RM_NGC5548} is
\begin{equation}
    \tau_0 \, c =
    \left( \frac{ X\, k\, \lambda_0 } { h\,c}
    \right)^{4/3}
    \left( \frac{ G \, M \,\dot{m}_{\rm Edd}} { 8 \, \pi\, \sigma} \, 
    \frac{ L_{\rm Edd}}
    {\eta \, c^2}
    \left( 3 + \kappa \right)
    \right)^{1/3} \ ,
\end{equation}
where $k$, $h$, $c$, $G$ and $\sigma$ are physical constants with their usual meanings, $X$ is a dimensionless parameter connecting wavelength with temperature, $\eta = L/\dot{M}\,c^2$ is the radiative efficiency of the accretion disc, and $\kappa$ is the ratio of external irradiative heating to internal viscous heating.
With conventional choices
$X=2.5$, $\eta=0.1$, and $\kappa=1$, for $\dot{m}_{\rm Edd}=0.15$,
our result for NGC\,4051 suggests an accretion disc size that is $\sim4$ times larger than expected.
Similarly, over-large disc sizes are reported in RM studies of other AGN \citep[e.g.,][]{Fausnaugh17, Cackett18, Edelson19, Netzer22, Guo2022, MM2023, Miller2025}. This ``larger-than-expected" disc size has also been observed in microlensing studies, where half-light radius measurements are found to be three times larger than standard disc predictions \citep[e.g.,][]{Morgan2010, Fian2021, MM2024}. Such a discrepancy implies that the standard accretion disc model may not fully describe the observed structure. Alternative scenarios, such as a slim/thick accretion disc or additional physical processes, including a magnetically coupled corona-disc system \citep{MSun2020_CHAR_model} or intrinsic temperature fluctuations in the disc \citep[e.g.,][]{Dexter_Agol2011, Ren2024} should be more broadly considered in future studies.

Allowing the power-law index $\beta$ to be a free parameter, we obtain a significantly better fit, yielding  $\beta = 3.48^{+0.35}_{-0.54}$  with $\chi^2_\nu = 2.38$. This more flexible model tries to capture the $u$-band excess by pushing $\beta$ to higher values and therefore inferring a smaller disc size $\tau_0 = 0.22^{+0.08}_{-0.04}$\,d. However, in both cases, the simple power-law model fails to fully capture the structure in the delay spectrum. Some alternative accretion disc models predict deviations from the standard temperature relation $T \propto R^{-3/4}.$ For example, in super-Eddington accretion discs, strong radiation pressure can significantly alter the temperature structure, leading to larger-than-expected lags. In this case, the accretion rate for NGC\,4051 is relatively modest ($\dot{m} \approx 0.2$; \citealt{Yuan2021}), making such an extreme scenario less likely. 


We also tested the physically motivated model from \citet{Kammoun2021,Kammoun2023}, which includes relativistic effects such as light-bending of the X-ray coronal irradiation close to the black hole within the lamp-post paradigm. 
The best fit is shown in Fig.~\ref{fig:delay_spectrum}
with parameters in Table\,\ref{tab:delay_spectrum_model}.
The main parameters are the accretion rate $\dot{m}_{\rm Edd}$, the corona height $h_{\rm c}$, and a colour correction factor $f_{\rm col}$ that boosts the temperature of the
reprocessed light above the local effective temperature:
$I_\nu = B_\nu(f_{\rm col}\,T_{\rm eff})/f_{\rm col}^4$ \ .
The predicted lags increase by raising any of these 3 parameters.
With the same procedure as described above, we fit the lag spectrum with this model by setting the colour-correction, $f_{\rm col}$, as a free parameter and fixed at 1.7 (as an average value). 
With the colour correction set free, we obtain $f_{\rm col}=2.4 \pm 0.3$, with accretion rate $\dot{m}_{\rm Edd} = 0.81\pm 0.05$ and corona height; $h_{\rm c}=16^{+29}_{-12ßß}$ $r_g$. 
Fixing $f_{\rm col}=1.7$ gives $\dot{m}_{\rm Edd} =0.92\pm 0.02$ and  $h_{\rm c} =17^{+19}_{-12}$ $r_g$. 
We find that for this model, the colour-correction $f_{\rm col}$ and corona height $h_{\rm c}$ are highly degenerate. 

The accretion rate inferred with the \citet{Kammoun2021} model is very high, close to the Eddington limit for both models. Previous studies have inferred low accretion rates \citep[e.g.,][]{Yuan2021} for NGC~4051 at odds with these results. Corroborating evidence does not suggest a high accretion rate (e.g., lack of strong \ion{Fe}{II} emission lines in the spectra). However, we note that these inferred accretion rates should be treated as upper limits, since the model assumes accretion disc reprocessing with no reprocessing
from larger regions. 

\begin{table}
\caption{{Rest-frame Delay Spectrum Models (Fig.\,\ref{fig:delay_spectrum}).
(a)~Power-law model. (b)~Lamp-post model of \citet{Kammoun2021}.}}
\label{tab:delay_spectrum_model}
    \centering
    \begin{tabular}{cccccc}
    \hline
    \multicolumn{5}{c}{(a) Power-law} \\
    \hline
       $\beta$  & $\tau_0/$day & $y_0$ & $\chi^2/{\rm dof}$ & $\chi^2_\nu$  
    \\ \hline 
        4/3  & $0.88\pm0.07$ & $0.92\pm0.03$ & 30.54/5 & 6.11
    \\ 
        $3.48^{+0.35}_{-0.54}$& $0.22^{+0.08}_{-0.04}$& $0.90 ^{+0.10}_{-0.12}$ & 9.53/4 & 2.38 
    \\ \hline
    \\
    \hline
    \multicolumn{5}{c}{(b) Lamp-post} \\
    \hline
         $f_{\rm col}$ & $\dot{m}_{\rm Edd}$ & $h_{\rm c}/r_g$ & $\chi^2/{\rm dof}$ & $\chi^2_\nu$
    \\ \hline 
        1.7 & $0.92\pm0.02$ & $17^{+19}_{-12}$ &  38.99/4 & 9.75 
    \\ 
        $2.4\pm0.3$ & $0.81\pm0.05$ & $16^{+29}_{-12}$ &  38.95/5 & 7.79
    \\ \hline
    \end{tabular}  
\end{table}

\subsubsection{Influence of the Diffuse Continuum Emission}


The results of this study provide compelling evidence for deviations from a simple power-law model for the observed lag spectrum. In particular, the $u$-band lag is 2.5$\sigma$ larger than power-law fits (see Fig.\,\ref{fig:delay_spectrum}). While this deviation may not be definitive on its own, it is certainly non-negligible and suggests additional physical processes beyond the standard lamp-post model. In the past, $u$-band excess lags have been observed in several sources, i.e., NGC\,5548 \citep{Edelson2015, Fausnaugh2016RM_NGC5548}, NGC\,4593 \citep{Cackett18}, a sample of five AGN \citep{Edelson19}, and recently in Fairall\,9 \citep{JVHS_F9_2020}, Mrk\,335 \citep{Lewin2023}.  More strikingly, our analysis of timescale-dependent lags (Fig\,\ref{fig:deltatrends}) reveals a significant increase in the $u$-band excess with lags for longer timescales, deviating by more than 3$\sigma$ from a power-law fit (for the model with $\beta$ = 4/3). 
The increasing excess on longer timescales is consistent with the presence of a distant reprocessing region contributing to the delayed emission (see Table~\ref{tab:timedelays-two_years_sim}), as seen in other AGN studies \citep[e.g.,][]{Donnan23_PG1119}.

One plausible interpretation is that this distant reprocessor could be associated with the BLR and/or an outflowing wind \citep[similarly seen for Mrk\,817,][]{Netzer_2024_Mrk817}. These structures naturally introduce longer delay components, as reprocessing from more distant regions imprints additional time lags in the observed light curves. The presence of such a reprocessor misaligns with the simplest reprocessing models and provides a strong argument that {Diffuse Continuum Emission (DCE)} \citep[free-free and free-bound recombination continuum,][]{KoristaGoad2001,Korista2019,Netzer2020} is not only present but also significant on long timescales \citep[similarly to Mrk\,335, NGC\,5548][]{DeRosa2015, Edelson2015, Cackett2022_freq_rs, Lewin2023}.

\subsection{Origin of the red AGN SED in NGC~4051}\label{sec:origin}

\begin{figure}
\centering
 \includegraphics[angle=-90,width=0.5\textwidth]{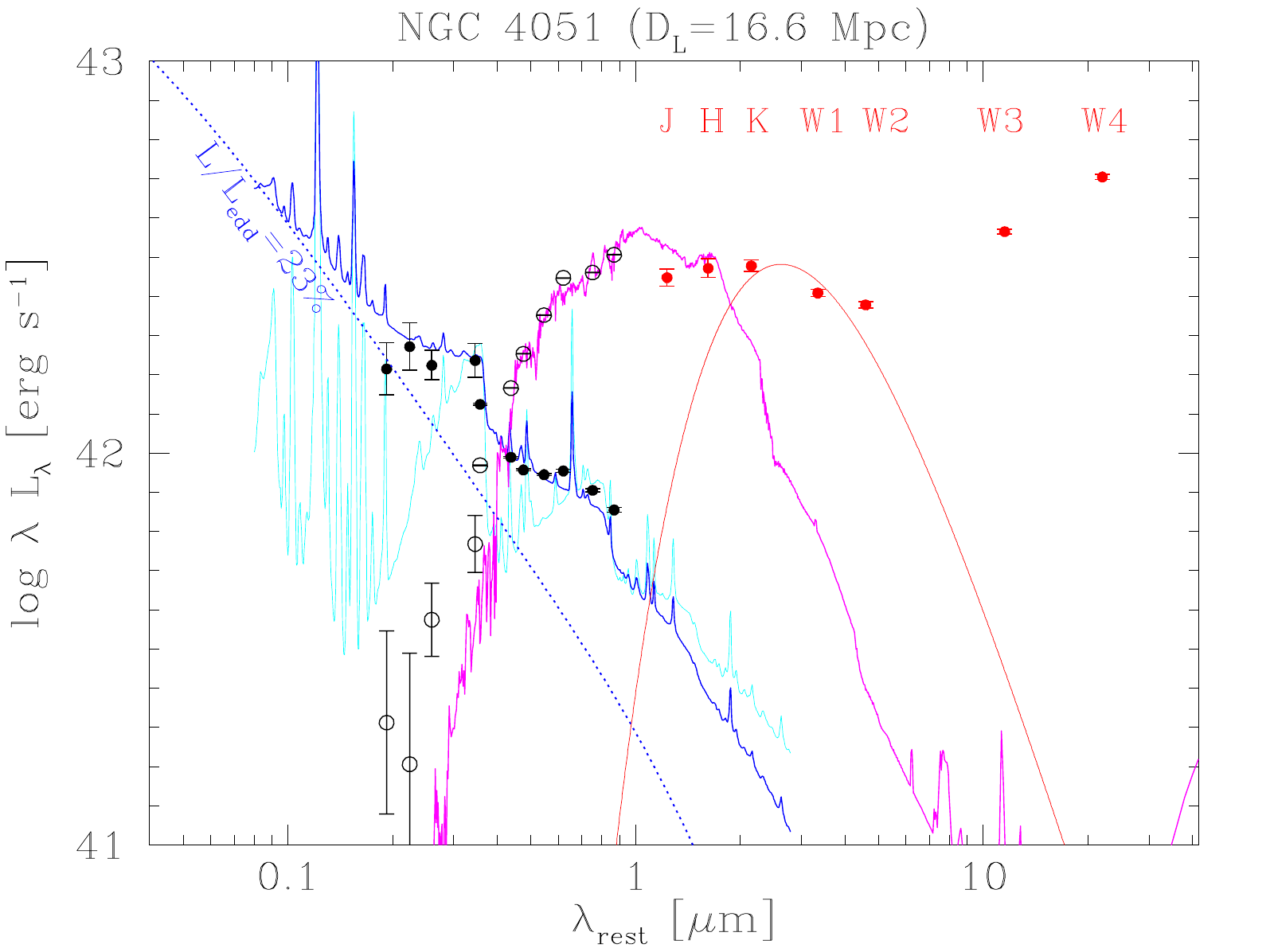}
    \caption{\label{fig:broadbandsed}
    The broad-band SED for NGC\,4051. The de-reddened galaxy (black open circles) and AGN (bright state, black filled circles) components from the flux-flux analysis are extended into the IR by the 2MASS and WISE photometry (red filled circles). The recovered galaxy spectrum is matched well by a spiral galaxy template (magenta), which also appears to dominate the 2MASS $J$ and $H$ bands. A blackbody with the average AGN hot dust temperature \citep[of $T \sim 1400$~K;][]{Landt2014} appears to be sampled by the 2MASS $K$, WISE W1 and W2 bands (red). The accretion disc$+$BLR DCE model from \citet{Netzer_2024_Mrk817} (blue) matches well the recovered AGN spectrum. The implied accretion disc has an Eddington ratio of $L/L_{\rm edd} \sim 23\%$, if an inclination angle of $i=0^\circ$ is assumed. The BLR DCE alone (cyan) dominates the LCO $u$ and $B$ bands (the Balmer continuum), as well as the LCO $r$, $i$, and $z$ bands (the Paschen continuum). 
}
\end{figure}  

Fig.~\ref{fig:broadbandsed} presents the broad-band SED for NGC\,4051 as it results from our Swift/LCO flux-flux analysis and extended into the infrared with non-simultaneous photometry from 2MASS and WISE. We show the bright-state AGN SED and the isolated constant component. The recovered spectrum of the constant component is well-matched by a spiral galaxy template \citep{swire}. As already found by \citet{Landt2013}, the host galaxy also seems to dominate the 2MASS $J$ and $H$ bands. The hot dust from the AGN torus has, on average, a temperature of $T_{\rm torus} \sim 1400$~K \citep{Landt2014} and a blackbody emissivity \citep{Landt2023b}. For NGC~4051, this component starts to dominate the emission in the 2MASS $K$ and WISE W1 and W2 bands (red solid line). A pure accretion disc spectrum does not seem to be a good approximation to the isolated AGN SED. Instead, the accretion disc$+$BLR diffuse continuum (DC) emission model presented by \citet{Netzer_2024_Mrk817} seems to match (blue solid line). Emission from the BLR DC alone (cyan solid line) seems to dominate the LCO $r$, $i$, and $z$ bands (the Paschen continuum) and the LCO $u$ and $B$ bands (the Balmer continuum). The implied Eddington ratio for the accretion disc in this combined model is $L/L_{\rm edd} \sim 23\%$.

\begin{figure}
\centering
\includegraphics[width=0.5\textwidth]{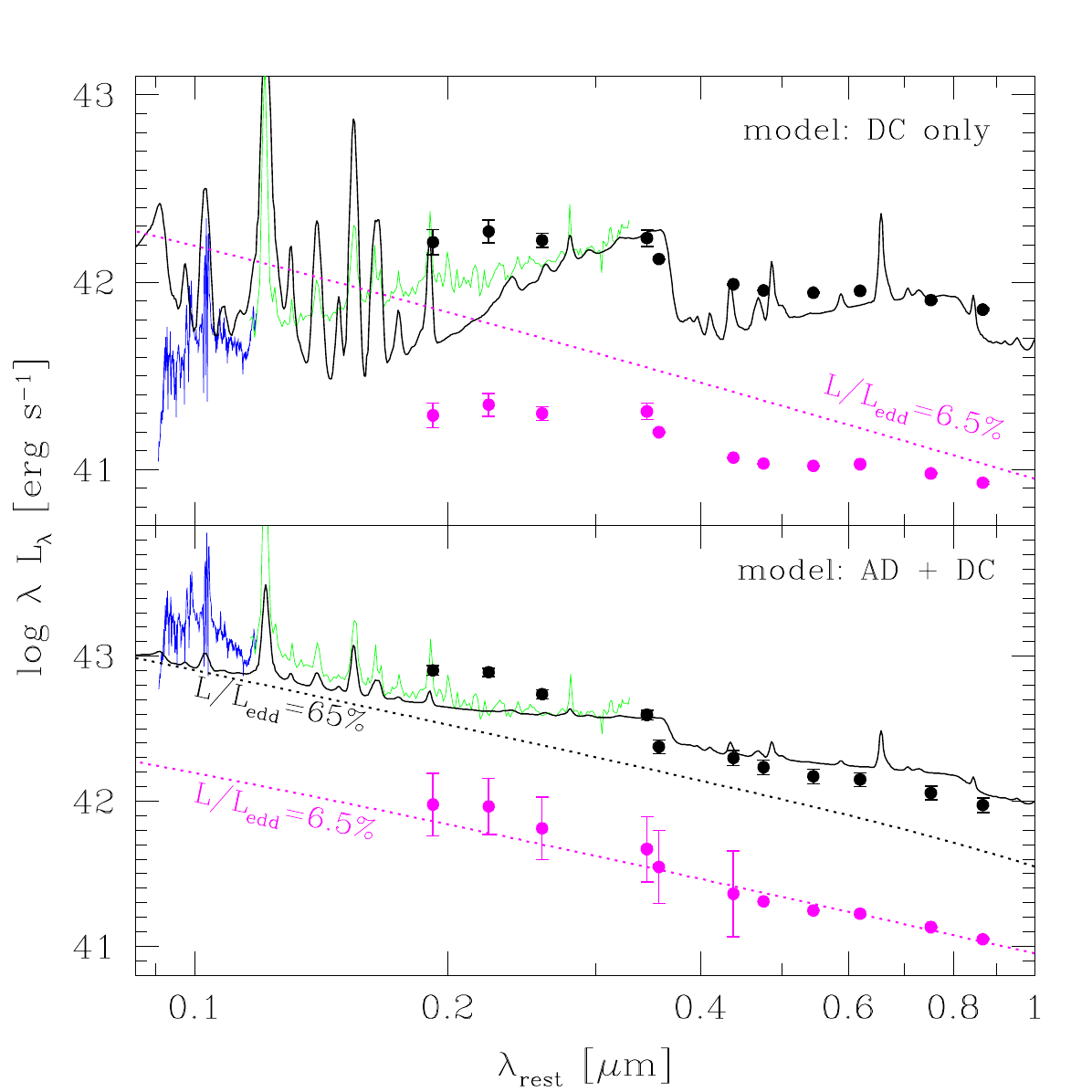}
    \caption{\label{fig:dustsed}
    The UV/optical AGN SED for NGC\,4051 as observed (top panel) and corrected for intrinsic dust extinction assuming a reddening of \mbox{E(B-V)=0.186} (bottom panel). The bright-state AGN (black filled circles) and faint-state AGN (magenta filled circles) isolated from the flux-flux analysis are shown together with the FUSE (blue) and IUE spectra (green) from \citet{Kaspi2004}. The BLR DC emission model from \citet{Netzer_2024_Mrk817} (black solid line, top panel) matches well the recovered AGN spectrum without the requirement of intrinsic reddening. On the other hand, the de-reddened AGN SED is well matched by the accretion disc$+$DC model from \citet{Netzer_2024_Mrk817} (black solid line, bottom panel). The implied accretion disc for the bright-state (black dotted line) has an Eddington ratio of $L/L_{\rm edd} \sim 65\%$, if an inclination angle of $i=0^\circ$ is assumed. On the other hand, the accretion disc with $L/L_{\rm edd} \sim 6.5\%$ (magenta dotted line), which illuminates the dusty torus, matches well the dust-corrected faint-state (bottom panel) but overestimates the observed (uncorrected) fluxes at wavelengths $\lambda_{\rm rest} \la 0.2~\mu$m (top panel). 
}
\end{figure} 

The combined accretion disc$+$BLR DC model predicts that at far-UV wavelengths, i.e., at wavelengths shorter than those probed by our \textit{Swift} observations, the accretion disc spectrum will dominate the emission. Therefore, we have supplemented our AGN SED with the \textit{FUSE} and \textit{IUE} spectra of \citet{Kaspi2004} and show it in Fig.~\ref{fig:dustsed} (top panel). Whereas the non-simultaneous far-UV spectra connect to the bright-state AGN \textit{Swift} fluxes, indicative of little variability between these epochs, the spectrum turns down instead of further rising as expected for an accretion disc. Emission from the BLR DC alone could explain the observed SED. But this option would imply very little emission from the accretion disc, which would be at odds with the observed X-ray flux (Fig. \ref{fig:Xray_spectrum}). Therefore, we considered the possibility that the relatively red AGN SED results from dust extinction.



For the additional (internal) dust, we assumed an SMC-like extinction law \citep{Gordon_SMC2003}
and applied it to the  $F_\nu\propto\nu^{1/3}$ power-law SED expected for a standard accretion disc \citep{SS73}. Fig.~\ref{fig:SED_Swift_LCO_b_f} shows the assumed power law SED for several $E(B-V)$ values. The best-fit model gives $E(B-V)=0.186\pm 0.025$, implying a factor of 5 extinction at 2000\,\AA.
Our best-fit value for the internal extinction is smaller than $E(B-V)= 0.37$ found by \citet{Jaffarian_Gaskell2020}, which, for a Milky Way-type extinction law, would produce a strong graphite feature at 2200~\AA. There is no evidence for this feature in the SED. 
Correcting the observed AGN SED for this internal extinction yields spectra consistent with the combined accretion disc$+$BLR DC emission model (Fig.~\ref{fig:dustsed}, bottom panel). The implied accretion rate for the bright-state AGN is $L/L_{\rm edd} \sim 65\%$, which is very similar to the accretion rate resulting from fitting the lag spectrum with only an accretion disc (see Fig.~\ref{fig:delay_spectrum}).

Already \citet{Kaspi2004} had considered the explanation that the relatively red far-UV spectrum is caused by internal reddening. However, they found this scenario to be unlikely since neither the X-ray spectrum nor the observed flux ratios between the UV and optical He II broad emission lines required additional extinction. Our best-fit value of $E(B-V) \sim 0.2$ is high enough that it should be detectable at soft X-ray energies even in X-ray spectra with modest count rates such as our simultaneous \textit{Swift} XRT spectrum. The \textit{Swift} spectrum does lack the soft X-ray excess usually observed to be strong in NGC~4051 \citep{Kaspi2004, Pounds2004} but can disappear at low-luminosity levels  \citep{Serafinelli2025ApJ}. Assuming the SMC gas-to-dust ratio, the estimated internal extinction translates to $N_{\rm H} = 1.9 \times 10^{22}$~cm$^{-2}$. This large hydrogen column would result in an obvious curvature in the soft X-rays, which has not been observed (with different observatories and at different luminosity levels). Therefore, if internal dust is required to explain the red spectrum, it would need to either reside in a gas-poor phase (as not to affect the X-rays) or have a specific line-of-sight to only affect the accretion disc light. We will further discuss the implications of this scenario in Sec.~\ref{interpretation}. 

\subsection{Origin of the anomalous lag spectrum}

\begin{figure}
\centering
\includegraphics[width=0.5\textwidth]{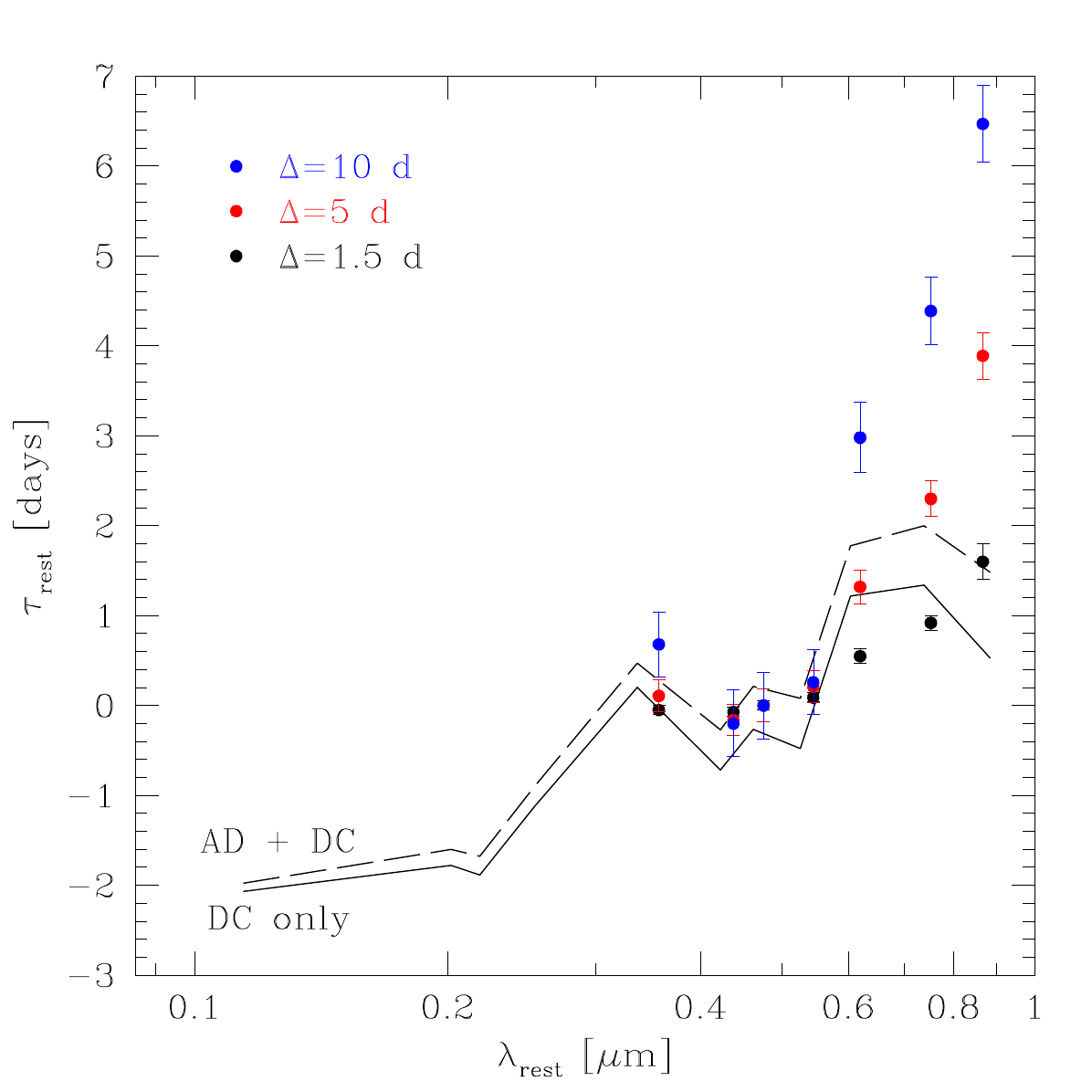}
    \caption{\label{fig:dustlag}
    The observed LCO UV/optical lag spectrum for NGC\,4051 (filled circles) compared to predictions by \citet{Netzer_2024_Mrk817} for the DC-only (black solid line) and AD$+$DC models (black dashed line) shown in Fig.~\ref{fig:dustsed}. Whereas both models match well the data at shorter wavelengths, none do so at the red wavelengths of the LCO $r$, $i$, and $z_s$ filters, which show a strong dependence on the {\sc PyROA} window function, $\Delta$. These red lags not only steeply increase with wavelength but also become larger as $\Delta$ increases from 1.5~d (black filled circles), to 5~d (red filled circles) and then to 10~d (blue filled circles).}
\end{figure}

Using an accretion disc, we showed in Section~\ref{discsize} that the lags in the reddest LCO bands ($r$, $i$, and $z$ bands) are much larger than expected from a standard reprocessing scenario. In Fig.~\ref{fig:dustlag}, we show the expected lag spectrum for the BLR DC model and the combined accretion disc$+$BLR DC model of \citet{Netzer_2024_Mrk817}. Similar to the pure accretion disc model, it cannot explain the lags in the $r$, $i$, and $z$ bands. The predicted shape in this spectral range is a slight rise between $r$ and $i$ followed by a decline between $i$ and $z$. But the red delays steeply rise and more so the larger the sampled variability scales (i.e., the larger the window function/smoothing). Therefore, it appears that a new AGN component is required by the observed delay spectrum, in addition to the accretion disc and BLR DC components.

\subsubsection{The Bowl model}\label{sec:bowlmodel}


The Bowl model, introduced by \citet{Starkey2023}, aims to fit simultaneously the observed lags and SED data across the UV and optical wavelength range as reverberations in the blackbody emission from a concave bowl-shaped disc surface that has a finite thickness $H$ with a power-law dependence on radius, $H \propto R^\beta$. For this purpose, it assumes the so-called lamp-post model, where a rapidly varying central X-ray source illuminates and heats an accretion disc with the disc having an intrinsic luminosity corresponding to the minimum AGN state. The luminosity of the disc to be heated is usually taken as that corresponding to the faint-state AGN resulting from the flux-flux analysis. The difference between the bright-state and faint-state AGN fluxes are then interpreted as the variable flux created by the heating process and giving rise to the observed lags. In other words, the Bowl model assumes that the (X-ray) lamp-post causes a faint disc to appear much brighter (by factors of a few). We will return to this important point in Sections~\ref{twodiscs}.

Our Bowl model fit for NGC\,4051 is presented in Fig.\,\ref{fig:bowl_ebmv}, where MCMC methods are used to sample the model parameters, constrained by the lag and SED data, and thus the geometry, temperature profiles, echo maps, lags, and SEDs predicted by the model.
{ The top panels of Fig.\,\ref{fig:bowl_ebmv} show the
AGN SED and lag data in comparison with the fitted model, in blue for the bright and in red for the faint state.
The lower panels show the corresponding temperature profile and disk geometry.
See the figure caption for details.}
This fit uses lag and SED data derived from the {\sc PyROA} fit with $\Delta=10$\,d, augmenting the LCO constraints with the \swift\ constraints on the SED
from the analysis in Section\,\ref{sec:swift}.
The data include the 7-band lag measurements $\tau(\lambda)$, and two disc SEDs: $F_\nu(\lambda)$ corresponding to the faintest and brightest states sampled during the campaign. 
Note that the constant host-galaxy component is subtracted so that we model just the reverberations from the disc reprocessing surface.

\subsubsection{Bowl Model Parameters}

{The Bowl model, in our current implementation, has 15 parameters, 
any of which can be fixed or varied to fit the lag and SED constraints.
The fit shown in Fig.\,\ref{fig:bowl_ebmv} has 8 fixed and 7 free parameters, leading to the}
parameter estimates and covariances shown in Fig.\,\ref{fig:bowl_ebmv_cov}, 
where $10^4$ MCMC samples trace the joint posterior distribution. 
The prior is uniform in $E(B-V)>0$, and uniform in the log of the other parameters.
We adopt $D_{\rm L}=16.6$\,Mpc, the Cepheid distance \citep{Yuan2021}, and a black hole mass $M_{\rm BH}=7.8\times10^5$\,M$_\odot$ \citep{BentzKatz2015}.
The Bowl geometry is specified by the disc inclination $i$, the disc radius $R_{\rm out}$, the opening angle at the outer rim $H/R$, and the power-law index $\beta$.
For this fit, we assume $i=45^\circ$.
The MCMC fit drives $\beta$ to large values to produce a flat inner disc with a steep outer rim, and we therefore fix $\beta=100$
{ as representative of these flat-disc and steep-rim geometries.}

The temperature profile on the flat inner disc falls as $T\propto R^{-3/4}$ in order to radiate the energy deposited in each disc annulus by viscous heating for an accretion rate $\dot{M}$ plus irradiation by two point sources, at height $H_x$ on the symmetry axis above and below the black hole, with bolometric luminosity $L_x$.
{ We fix the 
lamp-post height $H_x=10\,R_g$ and the disc inner radius $R_{\rm in}=6\,R_g$, the ISCO radius of a Schwarzschild black hole. 
A torque parameter fixed at $Q=0$ imposes the standard zero-torque boundary condition, so that viscous heating vanishes at $R_{\rm in}$.
We assume blackbody reprocessing, $f_{\rm col}=1$, rather than boosting colour temperatures of the reprocessed light.}

On a thin disc with $H< H_x\ll R$, viscous heating scales as $M\,\dot{M}/R^3$ and irradiation scales as $H_x\,L_x/R^3$.
{ As a result, } 
$\dot{M}$, $H_x$ and $L_x$ are degenerate parameters.
By convention, the faint disc SED fit assumes accretion only, thus setting an upper limit to $\dot{M}$. 
A brighter SED then arises from the change in irradiation, $H_x\,\Delta L_x$, which produces a multiplicative change in temperature, $\Delta T \propto T$, on the flat disc surface.
The change in irradiation produces a larger temperature change on the inward-facing slope of the steep outer-disc rim.

{ Finally, the MCMC fit adjusts two noise model parameters, $\sigma_0$ for the SED and $\sigma_\tau$ for the lags, which are added in quadrature with the nominal flux and lag uncertainties.
}

\subsubsection{Bowl Model fit to the inter-band lags}

Each element on the disc surface has a different time delay arising from light travel time being longer from the lamp to the reprocessing site to the observer than on the direct path from the lamp to { the} observer.
The delay distribution, or echo map, is produced by summing contributions from each surface element that represents the response to changes in the irradiation. The response at each surface element is weighted by the solid angle as seen from Earth and the change in blackbody surface brightness caused by a change in the lamp-post irradiation.
Summing over the surface elements gives the echo map, and the mean of { this} delay distribution gives the model { lag} prediction that we fit to the observed lags.

\begin{figure*}
\centering
    \includegraphics[width=\linewidth,trim=9mm 70mm 3cm 1cm,clip]{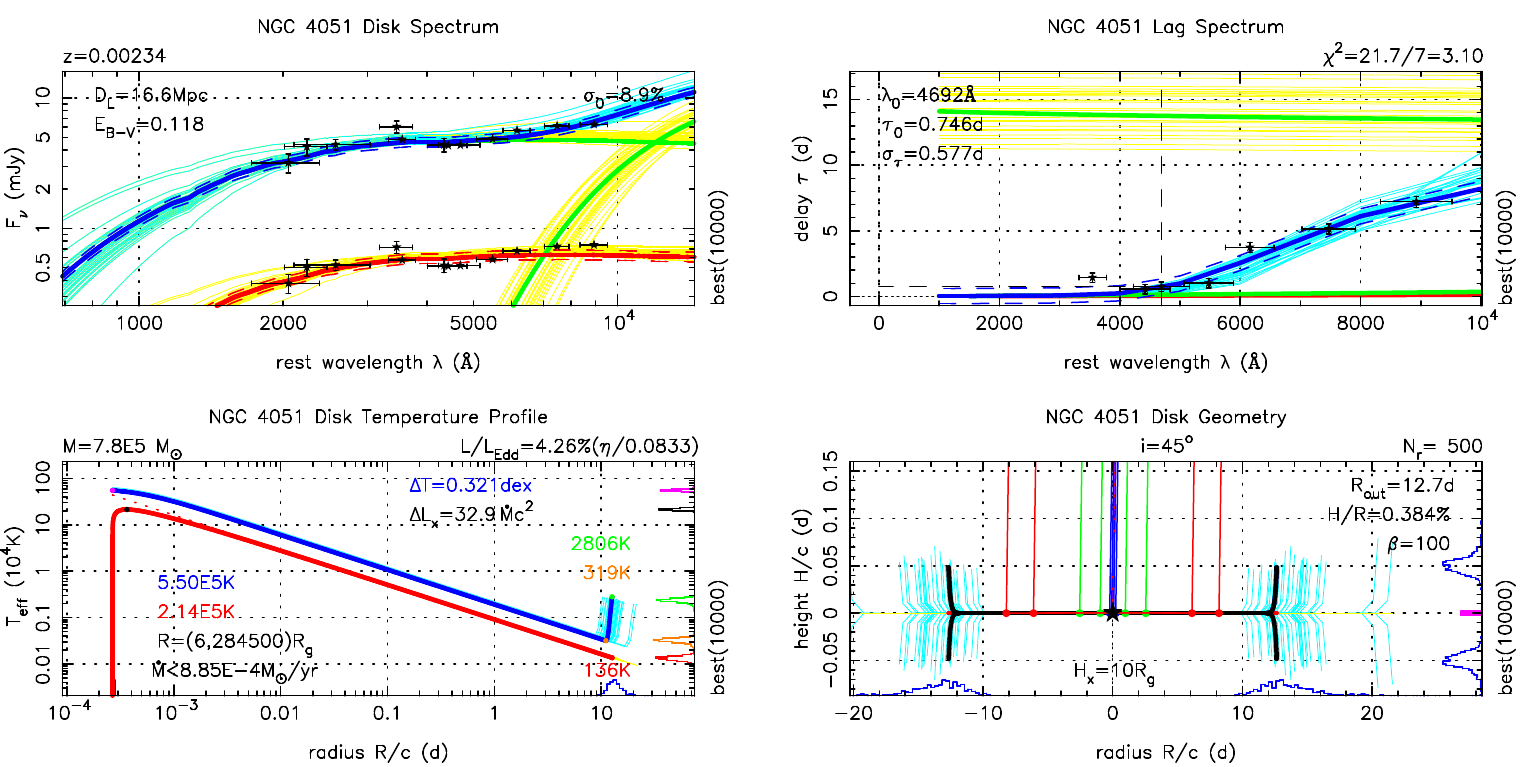}
    \caption{\label{fig:bowl_ebmv}
    Results of fitting the Bowl model simultaneously to the faint and bright AGN disc spectra (top left) and the PyROA lags for $\Delta=10$\,d (top right).
    The disc geometry (lower right) has a steep outer rim,
    resulting in the temperature  profile (lower left) falling as $T\propto R^{-3/4}$ and then rising on the outer rim.
    Red and blue curves correspond to the faint and bright state of the irradiated disc, respectively.
    Green curves in the top panels show the AGN broad-band spectrum and lag spectrum separately for the disc inside and outer edge outside the minimum temperature described in the model.
    A random selection of 30 MCMC samples (yellow and cyan) indicate uncertainties.
    Parameter values given on the plot are the best of $10^4$ MCMC samples (see Section\,\ref{sec:bowlmodel} for details and Fig.\,\ref{fig:bowl_ebmv_cov} for posterior parameter distributions).
    The faint temperature profile (red) corresponds to the accretion rate upper limit $\dot{M}$ for the black hole mass $M$ and Eddington ratio $L/L_{\rm Edd}$ indicated on the plot.
    The bright temperature profile (blue) is higher by $\Delta T$, corresponding to the indicated $\Delta L_{\rm x}$. 
    Coloured dots on the faint and bright temperature profiles mark the maximum and minimum temperatures, and the temperature at the rim, with similarly coloured median values and histograms on the right edge indicating uncertainties based on the MCMC samples.
    In the lower-right panel, coloured lines trace light rays from the lamp to the disc and up to the observer at $i=45^\circ$, and histograms on the right edge indicate uncertainties in the lamp height $H_{\rm x}$ and the rim height $H(R_{\rm out})$, in pink and blue respectively.
    Blue histograms on the lower edges 
    of the lower panels indicate uncertainty in the outer radius $R_{\rm out}$.
     }
\end{figure*}

Returning to Fig.\,\ref{fig:bowl_ebmv}, the Bowl model lags rise with wavelength, transitioning between the small lags (<1\,d) from the flat inner disc to the larger lag ($\sim8$\,d) from the steep outer rim. 
The model fits the observed lags well except for the $u$ band, which is several sigma above the model lag curve. The model predicts small lags (<1\,d) in the UV, where \swift\ data provide constraints on the SED but not on the lags.

We define the relatively flat inner disc and steep outer rim to be inside and outside the temperature minimum.
{ In the top panels of Fig\,\ref{fig:bowl_ebmv}, green curves with yellow uncertainty envelopes show the SED and lag predictions separately for these two regions.
The higher temperatures on the steep rim elevate the disc SED at wavelengths longer than 6000\,\AA.
The rim's increasing contribution at longer wavelengths then causes the model lag to rise, interpolating between the disc lag on the blue end toward the rim lag on the red end of the spectral range.
}

\subsubsection{Bowl Model fit to the SED}

The observed disc SED is approximately flat in $F_\nu$, turning down on the UV end.  
Our initial Bowl model fit (not shown) matches the observed disc SED provided the maximum disc temperature is $\sim3\times10^4$\,K.
Given the accretion rate set by the flux level and the relatively small black hole mass, the inner disc radius then needs to be $\sim100\,R_g$ to keep the inner-disc temperature low.
This is much larger than the innermost stable circular orbit (ISCO) radius of the black hole, requiring a large empty gap between the inner disc radius and the black hole.
The irradiated disc in this initial Bowl model then needs temperatures to rise by $\sim0.22$\,dex to match the observed rise in flux, and this makes the model disc SED somewhat too blue in the UV.
As these features are not entirely satisfactory, we opted to consider the possibility of significant dust extinction in the AGN host galaxy.

The Bowl model fit, shown in
Fig.\, \ref{fig:bowl_ebmv}, fits the observed disc SED by allowing for rest-frame SMC-like dust extinction to dim and redden the $F_\nu\propto \nu^{1/3}$ disc SED.
For the SMC-like dust extinction law, we interpolate among the SMC\,Average $A(\lambda)/A(V)$ from Table\,7 in \cite{Gordon2024}, with $R(V)\equiv A(V)/E(B-V)=3.02$.
{ By including SMC-like dust extinction and reddening with $E(B-V)\approx0.12$, the Bowl model with $R_{\rm in}=R_{\rm ISCO}$ succeeds in fitting both the observed SED and the inter-band lags.}



\subsubsection{A tale of two discs} \label{twodiscs}

The best-fit value of $R_{\rm out} = 13 \pm 3$~light-days for the location of the steep rim obtained by the Bowl model is within the range of $\sim 11-24$~light-days measured by the photometric NIR reverberation mapping campaign of \citet{Koshida2014} for the inner radius of the dusty torus in NGC~4051. Already, \citet{Breedt2010} suggested that the blackbody radiation of the hottest torus dust could explain some of the strong variability observed in the red part of the optical spectrum. But our Fig.~\ref{fig:broadbandsed} shows that the contribution of this component to the AGN $z$ band flux is negligible. Instead, a slightly hotter blackbody component is required to produce sufficient flux in the reddest LCO bands. The Bowl model estimates $T \sim 2800$~K for the maximum temperature of the steep rim.

Therefore, the new AGN component could be warm gas associated with the hot dust in the torus. But what heats these two co-spatial components to such different temperatures? Assuming radiative equilibrium and using eq. (3) of \citet{Landt2019}, we can estimate the UV/optical luminosity of the accretion disc that is illuminating and heating the dusty torus. For the average AGN hot dust temperature of $T=1400$~K \citep{Landt2014} and the lower limit on the dust radius posed by reverberation results of 11~light-days, this disc has an Eddington ratio of $L/L_{\rm Edd} = 6.5\%$. Such a low-luminosity disc is well reproduced by the dust-corrected faint-state AGN spectrum (see Fig.~\ref{fig:dustsed}, bottom panel) and is similar to the disc assumed to be irradiated and heated by the variable X-ray source in the Bowl model. On the other hand, using the same radiative equilibrium assumption, optically thick (dust-free) gas located at this dusty torus radius would generate the steep rim temperature that is required to explain the unusually large red lags only if it was illuminated by the UV/optical luminosity of a disc with $L/L_{\rm Edd} = 65\%$. This disc corresponds to the bright disc created by heating the faint disc in the Bowl model and is reproduced by the dust-corrected bright-state AGN spectrum in conjunction with emission from BLR DC (see Fig.~\ref{fig:dustsed}, bottom panel). 

\begin{figure*}
    \centering
\includegraphics[width=0.45\linewidth]
{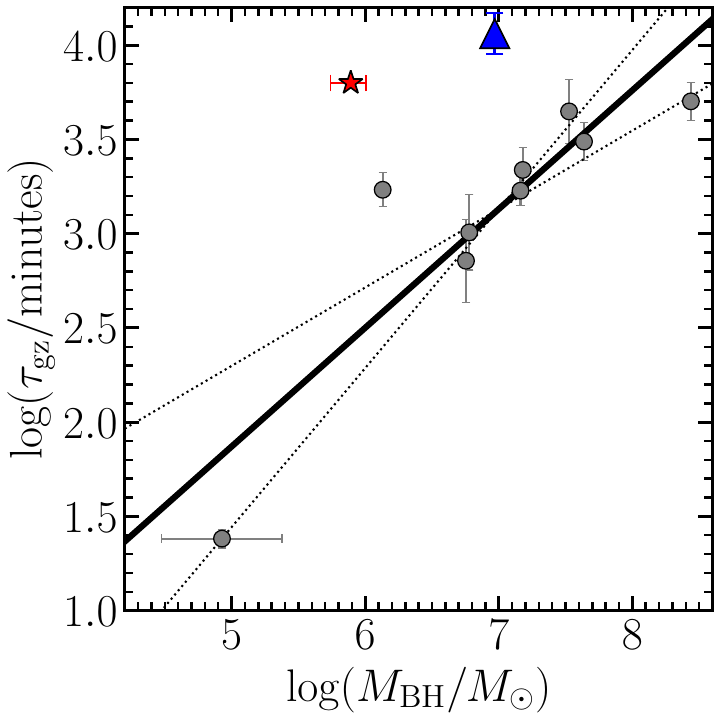}
\includegraphics[width=0.45\linewidth]
{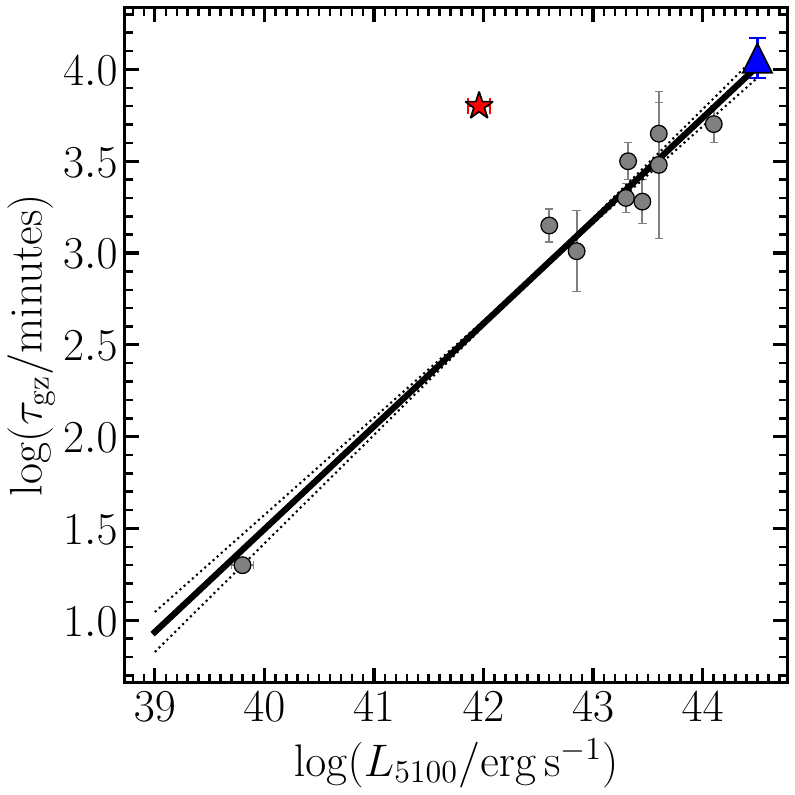}
\caption{\label{fig:NGC4051_montano}Inter-band lag from $g$ to $z_s$ as a function of AGN mass (left) and luminosity (right) \citet{Montano22} (grey; NGC\,4395 is the bottom left point). These scaling relations are particularly poorly sampled at their low ends, and thus the extrapolation may lead to bias in the relative contributions of the disc and BLR to the lag spectrum. NGC\,4051, marked by a red star, is a clear outlier. Lags were measured by {\sc PyCCF}, with the average value for both years to be consistent with the analysis in \citet{Montano22}. In blue, is another outlier --I Zw 1-- found by \citet{Drewes2026}, based on mass and luminosity found in \citet{Huang2019}.}
\end{figure*}

\subsection{Interpretation of the anomalous SED and lags} \label{interpretation}

We evaluate two competing scenarios to explain the unusually red UV/optical SED and the large inter-band time lags, synthesizing evidence from flux-flux analysis, time delays, and multi-wavelength spectroscopy.
\titlespacing*{\subsubsection}
  {0pt}{0.5\baselineskip}{0.3\baselineskip}
\subsubsection{Scenario 1: Intrinsic dust obscuration}
This model posits that the underlying emission is a standard accretion disc and BLR, reddened by extrinsic SMC-like dust.

\begin{itemize}
    \item \textbf{Consistency between reddened AGN SED and host galaxy:} 
    An SMC-type extinction law \citep{Gordon2024} applied to a standard disc (power-law) spectrum successfully reproduces the red variable component. Crucially, subtracting this model from the total light recovers a host galaxy SED which is not affected by this dust (orange curve, Fig. \ref{fig:SED_Swift_LCO_b_f}) that is in excellent agreement with independent flux-flux estimates. Therefore, the extinction is likely local to the AGN environment.
    
    \item \textbf{A low $L/L_{\text{Edd}}$ AGN is consistent with the X-ray emission:}
    The Bowl Model with intrinsic SMC-type dust extinction can successfully reproduce both the faint and bright AGN states, as well as the observed anomalous lag spectrum. This requires invoking a warm ($T \sim 2500$\,K) blackbody component, interpreted as optically thick, dust-free gas associated with the inner rim of the torus. The accretion disc in its faint state is powered by viscous heating only, with matter accreting at an inferred $L/L_{\text{Edd}} \sim 5\%$. The expected X-ray emission at this luminosity \citep{Kubota2018,Hagen2024} is consistent with observations. The bright state is thus powered by variable irradiation from the unobservable EUV region of the SED.
    

    \item \textbf{A low $N_H$ column is inconsistent with high dust content:} 
    The dust extinction required to fit the SED ($E(B-V)=0.186$) implies a neutral hydrogen column density of $N_H \approx 1.25 \times 10^{21}$\,cm$^{-2}$ assuming a normal gas-to-dust ratio \citep{Bohlin1978, Draine2003}. Integrating this column into the X-ray spectral fits produces strong residuals at $\sim 0.3$--$0.4$\,keV, contradicting high-resolution data \citep{Kaspi2004, Serafinelli2025ApJ} in which an absorbed combined powerlaw$+$blackbody component is sufficient to explain the continuum. To reconcile the optical reddening with the X-ray transparency, one must postulate an anomalous dust-to-gas ratio--specifically, a high dust content with negligible neutral gas affecting the accretion disc emission. This is opposite to the standard SMC-like dust/gas composition and implies a physical scenario where the line-of-sight is rich in dust but depleted of gas, or a geometric configuration where the compact X-ray corona is viewed through a clean line-of-sight while the larger UV/optical disc is obscured.

\end{itemize}

\subsubsection{Scenario 2: BLR-dominated emission model}
This model accepts the unobscured X-ray view and attributes the red SED to an intrinsic lack of strong UV disc emission, implying the observed flux is dominated by DCE from the BLR.

\begin{itemize}
    \item \textbf{Consistency with UV spectra:} 
    The combined \textit{FUSE/HST/IUE} spectra align better with the DC-only model than the dust-absorbed power-law (top panel, Fig. \ref{fig:dustsed}). The ``overblown'' $g-z$ lags (Fig. \ref{fig:NGC4051_montano}), which are anomalous for the source luminosity, are naturally explained if the optical response is dominated by an extended region (e.g., BLR and/or wind) rather than a compact accretion disc.

    \item \textbf{The energy budget challenge:} 
    While this model resolves the discrepant, obscuring columns inferred from X-ray and the dust scenario, it creates an energy balance paradox. To hide the accretion disc under the DC emission, the disc luminosity must be low ($< 5\%$ Eddington limit). However, the source exhibits strong X-ray luminosity ($L_X \gtrsim L_{UV}$).
     A standard thin disc that is faint enough to satisfy the UV constraints cannot provide sufficient seed photons to generate the observed X-rays. Consequently, this scenario explicitly requires a truncated disc or Advection-Dominated Accretion Flow (ADAF) in the inner regions to sustain the X-ray output. Furthermore, the radiative equilibrium analysis of the temperature of the dusty torus suggests heating from an accretion disc at $L/L_{\text{Edd}}\sim6.5\%$. The expected emission from this accretion disc in the FUV range is inconsistent (being higher than the observations) given the necessary luminosity to power the emission from the torus (see top panel of Fig.~\ref{fig:dustsed}).
\end{itemize}

In conclusion, the intrinsic dust model resolves the UV--optical SED shape but challenges the observed low absorption column inferred from X-ray data. This scenario requires a very high dust-to-gas ratio or a preferential line of sight where the X-ray emitting regions are not affected. In addition, it requires a warm gas component (dust free) located at the inner edge of the torus to boost the delays, particularly at the reddest optical bands. Conversely, the intrinsic BLR-only model adheres to the absorption column constraints in the X-rays and explains the anomalous lags, but requires a non-standard accretion flow (ADAF) to balance the X-ray/UV energy budget.

\section{Summary and Conclusion}
\label{sec:summary_conclusion}

We report the results of an intensive reverberation mapping campaign of NGC\,4051 using the Las Cumbres Observatory global robotic telescope network in seven optical bands, combined with archival \swift\ XRT and UVOT data. We apply several time-series analyses to measure the inter-band time delays and retrieve the SED of the variable AGN. Our main findings are summarised below:

\begin{itemize}

\item We find that inter-band delays increase with wavelength as $\tau \propto \lambda^{4/3}$, consistent with the results of numerous recent studies \citep[e.g.,][]{Cackett18, JVHS_F9_2020,Donnan23_PG1119}. The inferences from the normalisation (and thus the mass accretion rate) $\sim$ $\dot{m}_{\rm Edd} \simeq0.9$, depending on the assumptions made \citep{Cackett2007reprocessing_model,Kammoun2021,Starkey2023} (see Fig. \ref{fig:delay_spectrum}). The model implies accretion rates near the Eddington limit, much higher than earlier estimates \citep[e.g.,][]{Yuan2021}, though no spectral evidence (such as strong \ion{Fe}{II} lines) supports such high activity. Thus, values should be regarded as upper limits since the model assumes only accretion disc reprocessing.

\item  NGC\,4051 as many other AGN \citep[e.g.,][]{Fausnaugh2016RM_NGC5548,Edelson19,  Cackett18, Cackett20, JVHS_F9_2020} exhibit $u$-band excess, see Fig.\,\ref{fig:delay_spectrum}. When analysing the delay spectrum as a function of timescale with {\sc PyROA}, the $u$-band excess gets stronger for larger timescales (Fig. \ref{fig:deltatrends}). This suggests that the component which harbours the $u$-band excess is located further away from the central black hole.
\item The lightcurve analysis with {\sc MEMEcho} recovers asymmetric time delay distributions toward positive lags, with a prompt response (associated with the accretion disc) and a longer tail (seen in the redder bands, Fig.\,\ref{fig:Memecho}).
\item  A slow varying component with characteristic delay, $\Delta=10$\,d, is also required in the {\sc PyROA} to minimise the scatter in the residuals (in particular $r$, $i$ and $z_s$, Fig. \ref{fig:2year_wslow}). We propose the inner radius of the torus as a possible origin for this component \citep[e.g.,][]{Koshida2014,McHardy2018}. 

\item Consistent with theoretical predictions for the BLR \citep{KoristaGoad2001, Korista2019, Lawther2018, Netzer2020, Cackett2022_freq_rs}, we attribute the extended response to the DCE from the BLR, best evidenced by the excess lag in the $u$-band (see Fig.\,\ref{fig:deltatrends} and Fig.\,\ref{fig:delay_spectrum}). We introduce a diagnostic framework that systematically integrates lag-wavelength dependency with SED analysis, a combination that has been rarely utilized in previous studies
(Fig.\,\ref{fig:broadbandsed} and Fig.\,\ref{fig:SED_Swift_LCO_b_f}).

\item The unusually red UV/optical SED and extended time delays in NGC 4051 can be explained by two distinct physical models, each of which resolves specific observational challenges while creating others. The intrinsic dust model successfully recovers the host galaxy SED and spectral shape, but contradicts X-ray observations. This requires either a high dust-to-gas ratio or a specific viewing geometry to reconcile the high optical extinction with the low X-ray absorption column. Conversely, the BLR-dominated model is consistent with X-ray transparency and naturally explains the 'overblown' lags. However, it faces a severe energy budget paradox that cannot be resolved by a standard accretion disc.

\item Both scenarios necessitate physical components beyond the standard AGN paradigm. The dust model requires a discrete, dust-free gas component at the inner torus with a temperature of around 2500 K to drive the inter-band optical lags. The BLR-dominated model attributes the continuum to diffuse emission and thus implies that the underlying accretion disc is too faint to sustain the observed X-ray luminosity or torus heating. Consequently, if the red SED is intrinsic to the source, the central engine likely operates as a truncated disc or an ADAF in order to maintain an energy balance.

\item Ultimately, the data suggest a choice between anomalous dust properties (a high dust-to-gas ratio) and a non-standard accretion flow (an ADAF). Future simultaneous high-cadence UV/optical reverberation mapping combined with high-resolution X-ray spectroscopy is required to determine whether the variable emission originates from an obscured accretion disc or an intrinsically BLR-dominated continuum.

\end{itemize}

NGC\,4051 is one of the few supermassive black holes studied with high-cadence photometric observations with masses below $<10^6$~M$_\odot$  (the other two sources being NGC\,4395 \citep{Montano22, McHardy2023_NGC4395,Beard2025} and Pox~52 \citep{Sun2025}. These sources enable investigation of the poorly explored low-mass regime in scaling relations. NGC\,4051 is in fact an outlier in these relations \citep[see figure 2 in][]{Montano22}, with a $g-z_s$ lag of $\sim4.8$ days ($\log(\tau/{\rm min})\sim3.8$ as measured by {\sc PyCCF} \footnote{We used {\sc PyCCF} in order to be consistent with the analysis in \citep{Montano22}.} (average value for both years). This puts NGC\,4051 far above the expected 
extrapolation to low-masses ($\log(\tau/{\rm min})\sim2.4$, see Fig. \ref{fig:NGC4051_montano}). It is unclear why the large deviation is present in this particular source, but it may indicate that the BLR has far more influence on the measured lags than in other higher black hole mass AGN. 

Despite its comparatively low black hole mass, NGC\,4051 exhibits variability behaviour that closely follows the established wavelength-dependent lag trends observed in higher-mass Seyfert galaxies  \cite[e.g.,][]{Denney2010, Landt2013, Fausnaugh17, Edelson19}. This result extends the empirical validity of these relations to the lower end of the black hole mass distribution, demonstrating that the underlying accretion disc and reprocessing mechanisms remain operative in this regime. One of the key findings of this study is the growing significance of the $u$-band excess over longer timescales (see Table \ref{tab:timedelays-two_years_sim}). This suggests an increasing contribution from non-disc components or extended reprocessing regions. This behaviour provides an important constraint on models seeking to disentangle disc emission from line and continuum contamination, particularly in low-mass AGN where such effects may be amplified.

\section*{Acknowledgements}
We thank the anonymous referee for their detailed and thoughtful comments that have improved this manuscript.
This work makes use of observations from the Las Cumbres Observatory global telescope network. 
We acknowledge the use of public data from the \swift\ data archive. 
We thank Hagai Netzer for making his BLR DCE model available in electronic format.
We thank Rick Edelson for recommending NGC\,4051 for intensive monitoring with LCO.
M.V. gratefully acknowledges financial support from the Independent Research Fund Denmark via grant number DFF 3103-00146 and from the Carlsberg Foundation (grant CF23-0417).

\section*{Data Availability}
The raw datasets were derived from sources in the public domain: LCO archive \url{https://archive.lco.global} and \swift\ archive \url{https://www.swift.ac.uk/swift_live}. 
The inter-calibrated light curves are available in Zenodo, at~\href{https://zenodo.org/records/17242993}{10.5281/zenodo.17242993}.
This research made extensive use of {\sc astropy}, a community-developed core Python package for Astronomy \citep{Astropy-Collaboration:2013aa}, {\sc matplotlib} \citep{Hunter:2007aa} and {\sc corner} to visualize MCMC posterior distributions \citep{corner2016}.

\bibliography{bibliography.bib}
\bibliographystyle{mnras}

\appendix

\makeatletter
\renewcommand{\thesection}{\Alph{section}}
\renewcommand{\@seccntformat}[1]{Appendix~\csname the#1\endcsname:\quad}
\makeatother

\section{Cross-correlation Lag Measurements}\label{sec:pyccf}

\subsection{{\sc PyCCF} lag measurements}
\label{sec:pyccflags}

\begin{figure*}
\centering
    \includegraphics[width = 0.49 \textwidth]{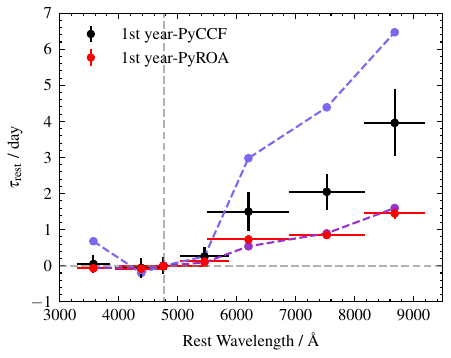}
    \includegraphics[width = 0.49 \textwidth]{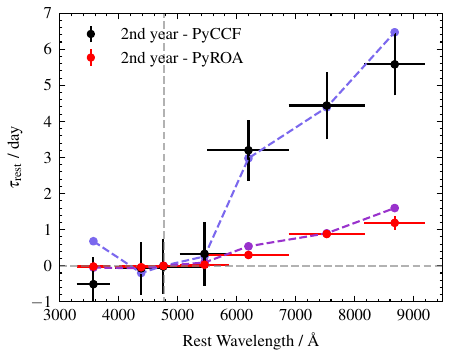}
    \caption{
    \label{fig:ROAvsCCF_dividedby_years}
       Comparison of centroid CCF lags (Table.\,\ref{tab:timedelays}) for Year\,1 (left) and Year\,2 (right) 
    as measired by {\sc PyCCF} (black)
    with the lags measured by {\sc PyROA} for $\Delta=1.5$\,d (red) for each year as discussed in Sec.~\ref{sec:yearbyyear_fits}.
   For reference, the dashed curves show {\sc PyROA} lags (see Table\,\ref{tab:timedelays-two_years_sim}) for fits to the full 2-year dataset for $\Delta=10$\,d (blue) and $\Delta=1.5$\,d (purple).
    Vertical and horizontal dashed grey lines are $g$-band as a reference wavelength and zero lag, respectively.
    While the {\sc PyROA} lags are consistent between the two years, the {\sc PyCCF} lags differ substantially (see text for details).
}
\end{figure*}

\renewcommand{\arraystretch}{1.5}
\begin{table*}
\caption{ \label{tab:timedelays}Observed-frame time lags relative to the $g$-band light curve. Measurement of the lags is divided by each year as measured by {\sc PyROA} and {\sc PyCCF}.}
    \begin{tabular}{c|ccccc|ccccc}
    \hline
    & \multicolumn{5}{c}{Year 1} & \multicolumn{5}{c}{Year 2}\\ 
    \hline
    Filter & $\tau_{\rm ROA}$  & $\tau_{\rm centroid}$ & $\tau_{\rm peak}$ & $r_{\rm max}$ & $F_{\mathrm{var}}$ &  $\tau_{\rm ROA}$  & $\tau_{\rm centroid}$ & $\tau_{\rm peak}$ & $r_{\rm max}$ & $F_{\mathrm{var}}$ \\
    & (day) & (day) & (day) & & & (day) & (day) & (day) & & \\
    \hline
    $\Delta$ & 1.5 & & & & & 1.5 & & & & \\
    \hline       
    $u$ & $-0.07 \pm 0.07$ & $0.05\pm 0.25$ & $0.15^{+0.25}_{-0.50}$ & $0.950$ & 0.093 &  $-0.02 \pm 0.09$ & $-0.49\pm0.75$ & $-0.10^{+0.55}_{-0.15}$ & $0.979$ & 0.170 \\
    $B$  & $-0.08 \pm 0.07$ & $-0.06\pm0.28$ & $-0.10^{+0.30}_{-0.20}$ & $0.958$ & 0.051 & $-0.04 \pm 0.08$ & $-0.10\pm0.78$ & $0.20^{+0.10}_{-0.40}$ & $0.985$ & 0.099\\
    $g$  & $0.00 \pm 0.07$ & $0.00\pm0.23$ & $0.00^{+0.00}_{-0.00}$ & $1.0$  & 0.039 & $0.00 \pm 0.08$ & $-0.01\pm0.70$ & $0.00^{+0.00}_{-0.00}$ & $1.0$ & 0.081 \\
    $V$  & $0.12 \pm 0.07$ & $0.25\pm0.28$ & $-0.10^{+0.40}_{-0.15}$ & $0.949$ & 0.032 & $0.03 \pm 0.09$ & $0.37\pm0.90$ & $0.25^{+0.10}_{-0.45}$ & $0.982$ & 0.063 \\
    $r$  & $0.73 \pm 0.11$ & $1.51\pm0.57$ & $0.45^{+0.15}_{-0.35}$ & $0.903$ & 0.029 & $0.30 \pm 0.12$ & $3.17\pm0.92$ & $0.35^{+0.15}_{-0.60}$ & $0.957$ & 0.052\\
    $i$  & $0.85 \pm 0.10$ & $2.07\pm0.53$ & $0.45^{+0.10}_{-0.45}$ & $0.891$ & 0.030 &  $0.88 \pm 0.12$ & $4.46\pm0.94$ & $0.45^{+0.30}_{-0.10}$ & $0.958$ & 0.045\\
    $z_s$  & $1.45 \pm 0.14$ & $3.94\pm0.90$ & $1.35^{+0.70}_{-0.90}$ & $0.838$ & 0.031 & $1.19 \pm 0.19$ & $5.62\pm0.84$ & $0.55^{+1.80}_{-0.20}$ & $0.919$ & 0.040 \\
    \hline     
    \end{tabular}   
\end{table*}


A traditional method of measuring inter-band lags relies on the CCF between pairs of light curves. In this Appendix, we present CCF lag measurements for comparison with those obtained with {\sc PyROA}.

The CCF method computes the correlation coefficient $r(\tau)$ between one light curve and another for a grid of time shifts $\tau$, selecting either the peak lag $\tau_{\rm peak}$ that maximises $r(\tau)$ 
or the centroid lag $\tau_{\rm centroid}$, a weighted average of values near the peak.
Standard practice for AGN
centroid lags is a weighted mean of lags with $r(\tau) > 0.8\,r_{\rm max}$, where $r_{\rm max}$ is the maximum correlation coefficient \citep{Peterson04}.

To measure CCF lags, we used the {\sc PyCCF} code \citep{PyCCF_MSun_2018}, which implements the interpolated CCF (ICCF) algorithm \citep{Gaskell1987, Peterson1998PASP}. 
As AGN light curves are typically unevenly sampled in time, the ICCF algorithm interpolates linearly between the epochs in one light curve to match those in the other when shifted by $\tau$.
The process is then reversed, and the results are averaged. 
Thus, both light curves are interpolated, and the auto-correlation function of one light curve with itself is symmetric.
We adopt the $g$-band light curve as the reference, as it has the highest signal-to-noise ratio. 
To avoid possible biases
arising from linear interpolation across the seasonal gap, we opted to measure CCF lags separately for Year\,1 and Year\,2. 
The lag uncertainties are estimated by repeating the entire process $10^4$ times on mock light curves obtained with the flux randomization/random subset selection (FR/RSS) method
\citep{Peterson1998PASP}.  

Table~\ref{tab:timedelays} collects
the measured CCF lags, both $\tau_{\rm peak}$ and $\tau_{\rm centroid}$, and corresponding
$r_{\rm max}$. 
The peak correlation coefficients
$r_\mathrm{max}$ indicate strongly correlated variations across all bands, as is evident from the light curves.
The CCF lags for both years exhibit the same general trend of lags increasing with wavelength.
This trend is as expected for a central lamp-post irradiating an accretion disc, due to the disc temperature decreasing with radius \citep[e.g.,][]{Cackett2007reprocessing_model} and is consistent with many previous AGN inter-band lag measurements \citep[e.g.,][]{Fausnaugh2016RM_NGC5548,Edelson19}.

The CCF centroid lags $\tau_{\rm cent}$ are typically 3 times larger than the CCF peak lags $\tau_{\rm peak}$
in Year\,1, and up to 10 times larger in Year\,2.
The pattern of $\tau_{\rm peak} < \tau_{\rm centroid}$ may
be interpreted as evidence for an asymmetric delay distribution with positive skew. 
This can arise, for example, because responses at negative lags violate causality, and a long response tail with positive lags can arise from reprocessing over an extended region.

The CCF centroid lags $\tau_{\rm cent}$ for $r$, $i$, and $z_s$ are larger by a factor of two in Year\,2 than in Year\,1, albeit at $<2\sigma$. 
This apparent change may occur because CCF lags are sensitive not only to the delay distribution of the response, but also to the character of the light curve that drives the delayed responses.
A larger amplitude of slower variations in Year\,2 compared with Year\,1 may thus account for this apparent change in the CCF lags.
Note that in contrast to the CCF lags, the ROA lag measurements do not change significantly between Years\,1 and 2, suggesting that the delay distribution has not changed.
These results raise a cautionary note that changes in CCF lags do not necessarily imply changes in the size of the reprocessing region.

\begin{figure*}
\centering
\includegraphics[width=0.9\textwidth]
{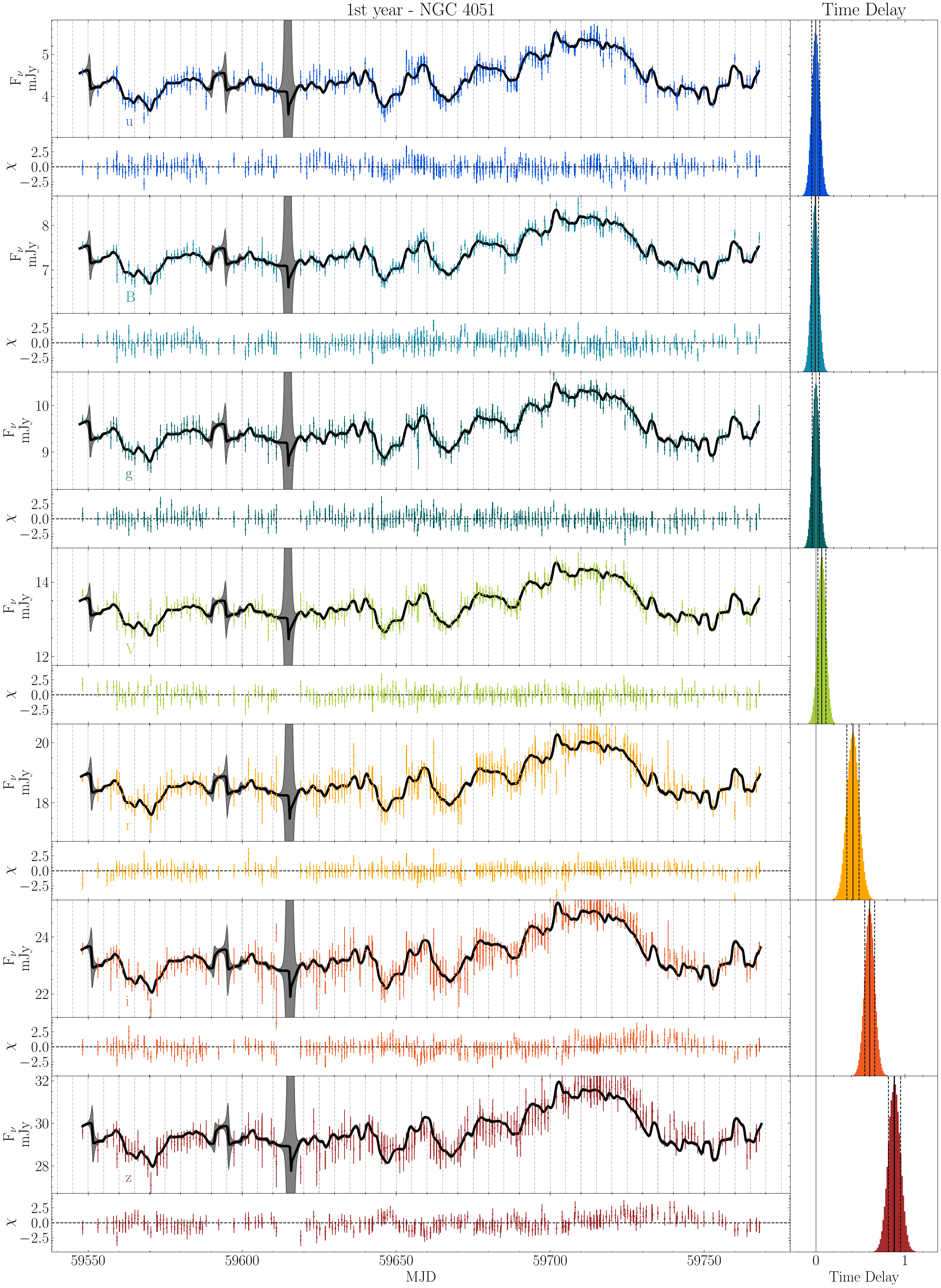}
\caption{
\label{fig:roa_lc_each_year_1styear} Light curves of NGC\,4051 in Year\,1 of our 7-band LCO monitoring campaign.
In each panel, the best-fit {\sc PyROA} lightcurve model is overlaid in black with a grey uncertainty envelope. 
This PyROA fit minimises the BIC for $\Delta=0.65$\,d, giving a very flexible light curve shape $X(t)$. 
The grey uncertainty envelope expands in data gaps of just a few days.
The lag measurements increasing with wavelength, as shown in the right column, are detailed in Table\, \ref{tab:lag_1_2_d065_046}.
}
\end{figure*}

\begin{figure*}
\centering
\includegraphics[width=0.9\textwidth]{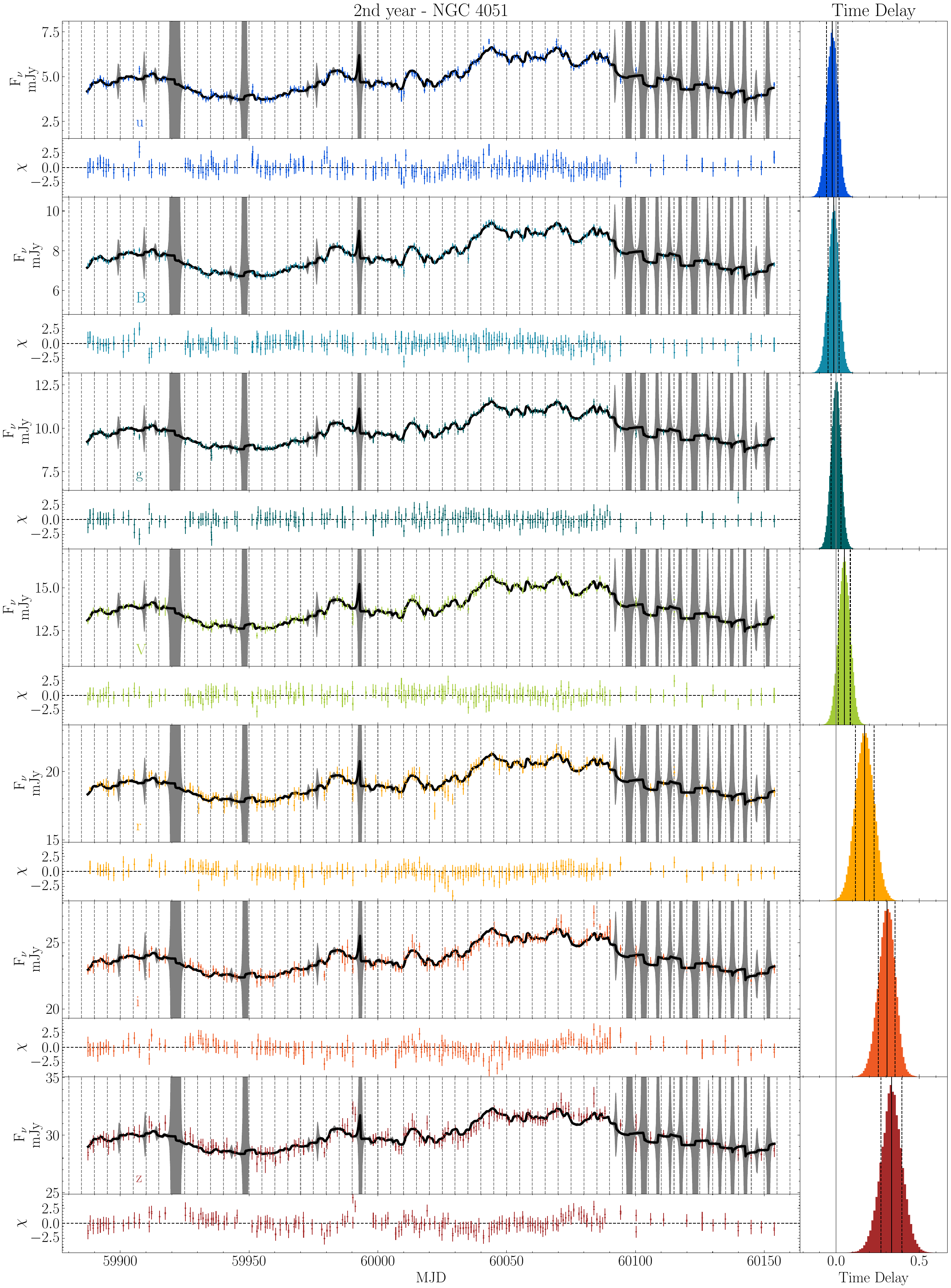}
\caption{
\label{fig:roa_lc_each_year_2ndyear}
Light curves of NGC\,4051 in Year\,2 of our 7-band LCO monitoring campaign.
In each panel, the best-fit {\sc PyROA} lightcurve model is overlaid in black with a grey uncertainty envelope. 
This PyROA fit minimises the BIC for $\Delta=0.46$\,d, giving a very flexible light curve shape $X(t)$. 
The uncertainty envelope expands greatly in data gaps of just a few days.
The lag measurements increasing with wavelength, as shown in the right column, are detailed in Table\, \ref{tab:lag_1_2_d065_046}.
}
\end{figure*} 

\begin{table}
\centering
\caption{ 
\label{tab:lag_1_2_d065_046}
Observed-frame time lags relative to the $g$-band light curve measured by {\sc PyROA} fits to the LCO data in the individual Years\,1 and 2.  The
BIC is minimised for $\Delta$ = 0.65\,d in Year\,1 and 0.46\,d in Year\,2.}
\begin{tabular}{l|cc}
\hline
 & Year 1 & Year 2 \\
\hline
Filter& $\tau_{\rm PyROA}$ & $\tau_{\rm PyROA}$ \\
& (day) & (day)\\
\hline
$\Delta$ & $0.65\pm0.02$ & $0.46\pm0.02$ \\
\hline
$u$ & $0.00\pm0.05$ & $-0.02\pm0.03$ \\
$B$ & $-0.01\pm0.04$ & $-0.01\pm0.03$ \\
$g$ & $0.00\pm0.04$ &  $0.00\pm0.03$ \\
$V$ & $0.07\pm0.05$ & $0.05\pm0.04$ \\
$r$ & $0.42\pm0.07$ & $0.17\pm0.06$ \\
$i$ & $0.60\pm0.06$ & $0.31\pm0.05$ \\
$z_s$ & $0.88\pm0.07$ & $0.33\pm0.06$ \\
\hline
\end{tabular}
\end{table}

\subsection{Comparison of 
{\sc PyCCF} and {\sc PyROA} lags }
\label{sec:yearbyyear_fits}

In order to directly compare the {\sc PyROA} and {\sc PyCCF} lag measurements, we performed {\sc PyROA} fits for the Year\,1 and Year\,2 LCO monitoring datasets separately.
The resulting fits to the light curves are shown in Figs.\,\ref{fig:roa_lc_each_year_1styear} and \ref{fig:roa_lc_each_year_2ndyear}, 
and the lags are given in Table~\ref{tab:timedelays}.

The delay spectra from both methods, CCF and ROA, are shown in Fig.\,\ref{fig:ROAvsCCF_dividedby_years} for a direct comparison (see also Table\,\ref{tab:timedelays}). The general trend is the same for both methods, where we find an increase in the time delay with the increase of the wavelength. However, we find that the CCF produces larger lags than ROA, a trend observed in similar studies \citep{Donnan23_PG1119, Prince2025}
This is likely due to the asymmetry in the transfer function, which has long tails at large delays \citep{Starkey2016,Cackett2022_freq_rs}. Thus, the CCF will tend to flatten out at positive delays from the peak, displacing the centroid of the CCF for correlation coefficients $\leq0.8\,r_{\rm max}$. Furthermore, {\sc PyROA} tends to find solutions for small $\Delta$ values, which concentrates the lag towards smaller values as the fast reverberation signals will be concentrated in these high-frequency features. The focus of each method on different aspects of the light curves produces this natural discrepancy, which gets accentuated at longer wavelengths. {\sc PyROA} finds uncertainties $\sim5$ times smaller than {\sc PyCCF}, as demonstrated by simulations \citep{pyroa21}.

\subsection{Variability amplitude}

Table~\ref{tab:timedelays} also lists for each year the lightcurve fractional rms variability amplitude
\citep{Vaughan2003}:
\begin{equation}
    F_{\mathrm{var}} \equiv \frac{\sqrt{S^2-\bar\sigma^2}}{\bar{F}} \ ,
\end{equation}
where $\bar{F}$ and $S^2$ denote the sample mean and variance of the flux data, and $\bar\sigma^2$ is the mean 
of the squared flux uncertainties, subtracted to remove the noise contribution to $S^2$.
Note that $F_{\rm var}$ decreases with wavelength, from 9\% and 17\% in the $u$ band in Years\,1 and 2, respectively,
to 3\% and 4\% in the $z_s$ band.
This trend reflects the relatively blue SED of the variable (AGN disc) component
being diluted by the relatively red SED of the constant (host galaxy) component of the AGN light.

\section{ Bowl Model Parameter Covariances}
\label{sec:bowlcov}

\begin{figure*}
\centering
    \includegraphics[width=\linewidth]{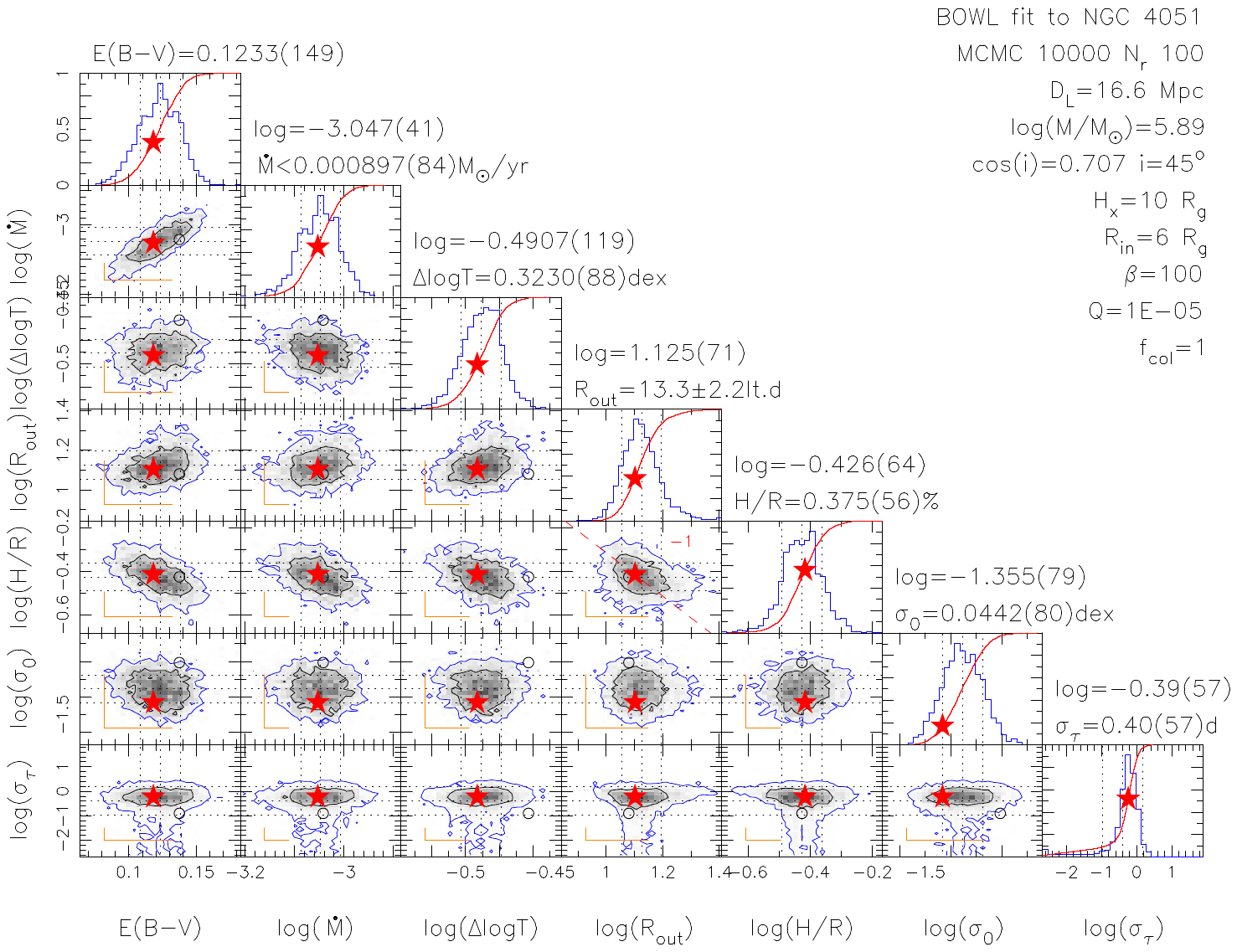}
\caption{\label{fig:bowl_ebmv_cov}Parameter posteriors, defined by $10^4$ MCMC samples, for the Bowl model fit in Fig.\,\ref{fig:bowl_ebmv}.
    The prior is uniform for $E(B-V)$, and uniform in the log of the other fit parameters.
    Note the positive correlation between $E(B-V)$ and accretion rate $\dot{M}$.
     }
\end{figure*}

Fig.\,\ref{fig:bowl_ebmv_cov} shows the Bowl model parameter covariances based on $10^4$ MCMC samples.

\end{document}